\documentclass[sigconf]{acmart}
\usepackage{multirow}
\usepackage{array}
\usepackage{algorithm}
\usepackage{algpseudocode}
\usepackage{xspace}
\usepackage{soul}
\usepackage{graphicx}
\usepackage{etoolbox}
\usepackage{float}
\usepackage[utf8]{inputenc}
\usepackage{xcolor}
\usepackage{balance}  
\usepackage{listings}
\usepackage{hyperref}

\usepackage{tabularx}
\usepackage{caption}
\usepackage{mathtools}
\usepackage{url}
\usepackage{amsfonts}
\usepackage{textcomp}

\usepackage{color}

\usepackage{subcaption}
\usepackage{enumitem}

\definecolor{quotegray}{gray}{0.9}
\usepackage{threeparttable}

\definecolor{lightgreen}{RGB}{129,199,132} 
\definecolor{darkgreen}{RGB}{56,142,60}    

\usepackage[most]{tcolorbox}
\tcbset{colback=gray!15, colframe=white, boxrule=0pt, arc=2pt, left=4pt, right=4pt, top=2pt, bottom=2pt}

\usepackage{soul}

\sethlcolor{gray!20}

\newcolumntype{L}[1]{>{\raggedright\let\newline\\\arraybackslash\hspace{0pt}}m{#1}}
\newcolumntype{C}[1]{>{\centering\let\newline\\\arraybackslash\hspace{0pt}}m{#1}}
\newcolumntype{R}[1]{>{\raggedleft\let\newline\\\arraybackslash\hspace{0pt}}m{#1}}

\usepackage[normalem]{ulem}

\definecolor{improved_green}{HTML}{49D149}
  
\definecolor{worsened_red}{RGB}{255,77,61}  
\definecolor{drd2blue}{RGB}{0,153,255}      
\definecolor{darkgreen}{RGB}{56,142,60}
\definecolor{blue1}{RGB}{26,35,126}   
\definecolor{blue2}{RGB}{66,165,245}  
\definecolor{blue3}{RGB}{144,202,249} 

\usepackage{booktabs}
\usepackage{amssymb} 

\usepackage[normalem]{ulem}

\AtBeginDocument{%
  }

\setcopyright{acmlicensed}
\copyrightyear{2025}
\acmYear{2025}
\acmDOI{XXXXXXX.XXXXXXX}

\acmConference[Conference acronym 'XX]{Make sure to enter the correct
  conference title from your rights confirmation emai}{June 03--05,
  2025}{Woodstock, NY}
\acmISBN{978-1-4503-XXXX-X/18/06}

\newcommand{\system}{\emph{HALO}\xspace}
\newcommand{\fone}{\emph{MolCluster}\xspace}
\newcommand{\ftwo}{\emph{MolStrategy}\xspace}
\newcommand{\fthree}{\emph{MolSynthesis}\xspace}

\makeatletter

\newcommand{\RomanNum}[1]{\uppercase\expandafter{\romannumeral #1}}

\begin{document}


\title{\system: Interactive Co-abductive Reasoning in Scientific Hypothesis Generation}

\author{Youngseung Jeon}
\email{ysj@ucla.edu}
\affiliation{%
  \institution{University of California}
  \city{Los Angeles, CA}
  \country{USA}
}

\author{Kat Limqueco}
\email{katlimq@ucla.edu}
\affiliation{%
  \institution{University of California}
  \city{Los Angeles, CA}
  \country{USA}
}

\author{JiaSyuan Chang}
\email{serenacjs@ucla.edu}
\affiliation{%
  \institution{University of California}
  \city{Los Angeles, CA}
  \country{USA}
}

\author{Xiang `Anthony' Chen}
\email{xac@ucla.edu}
\affiliation{%
  \institution{University of California}
  \city{Los Angeles, CA}
  \country{USA}
}

\renewcommand{\shortauthors}{Jeon et al.}

\begin{abstract}
Scientific discovery is essential yet inefficient, primarily because generating hypotheses within a vast search space hinders breakthroughs.
While current AI systems assist in generating new hypothesis candidates, they lack interactive support for the reasoning process by which users develop these outputs into promising hypotheses, resulting in surface-level hypotheses.
To address this issue, we present a co-abduction, a human-AI collaborative framework for abductive reasoning in scientific hypothesis generation.
To operationalize co-abduction, we build \system, a human–AI collaborative system for molecular hypothesis generation in drug discovery, enabling improved candidate clustering, strategy identification, and multi-strategy synthesis. 
In expert studies involving 10 medicinal chemists, \system significantly facilitated abductive reasoning for hypothesis generation—efficient candidate observation, systematic strategy identification, and coherent multi-strategy composition—and enabled participants to produce higher-quality, more diverse candidate molecules.

\end{abstract}

\begin{teaserfigure}
        \centering
    \includegraphics[width=0.95\textwidth]{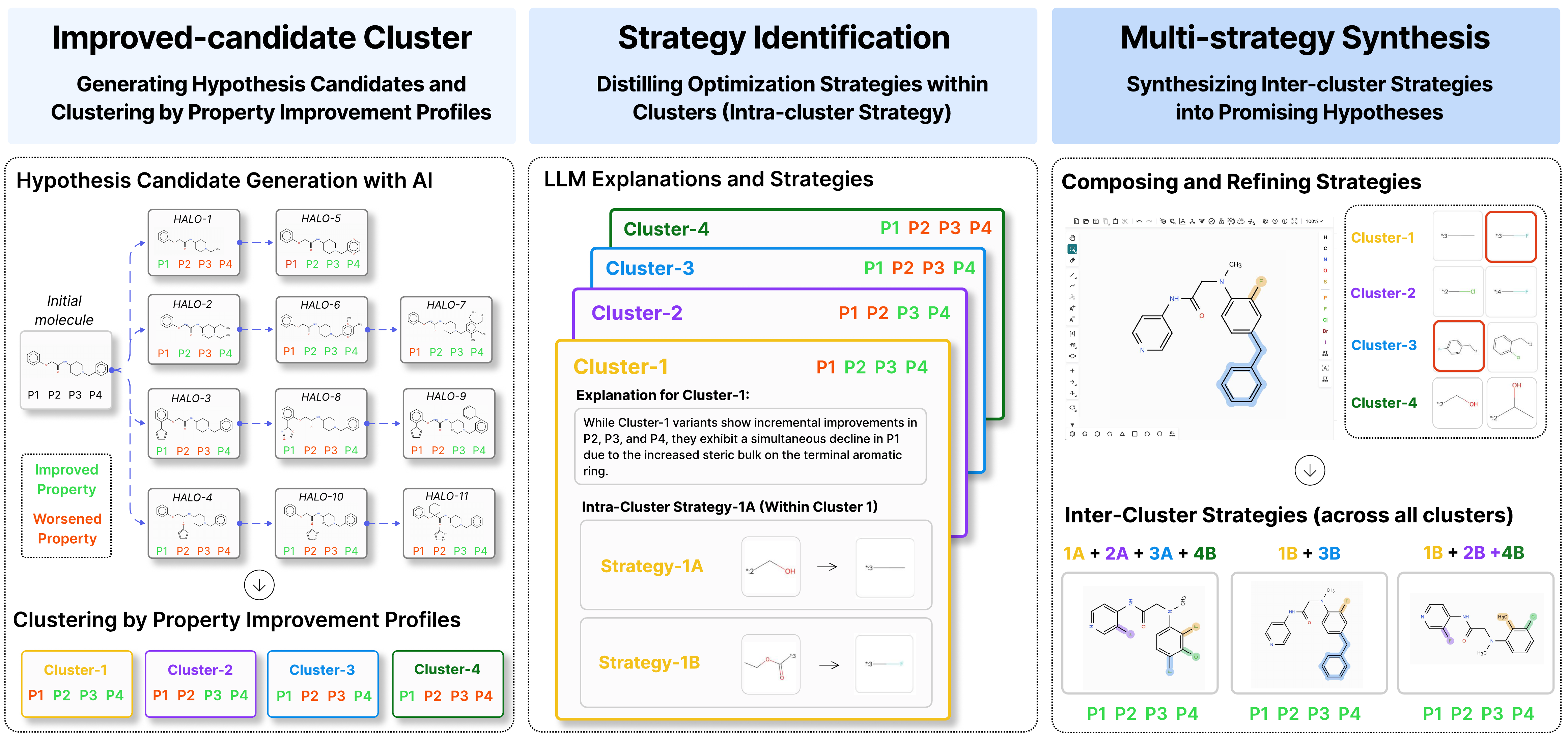}
    \caption{
    \system is a human-AI collaborative system that assists medicinal chemists in abductive reasoning to generate promising molecular hypotheses in drug discovery. 
    \system supports abductive reasoning through three stages: (1) Improved-candidate Cluster, experts and AI generate molecular candidates and cluster them by property improvement profiles (\textcolor{improved_green}{green: augmented}, \textcolor{worsened_red}{red: worsened}); (2) Strategy Identification, a large language model (LLM) generates explanations for clusters and provides intra-cluster strategies to improve the worsened property; 3) Multi-strategy Synthesis, LLM suggests combinations of multiple strategies across clusters (inter-cluster strategies) and experts iteratively synthesize them until targeted properties are satisfied.}
    \label{fig:first}
\end{teaserfigure}




\maketitle
\makeatother

\section{Introduction}

Scientific discovery advances human progress by generating new knowledge, entities, and mechanisms~\cite{simon2019sciences}.
Penicillin saved millions of lives by curing previously lethal infections~\cite{drews2000drug}, and semiconductors enabled modern computing~\cite{moore1965cramming}.
Achieving these breakthroughs required researchers to identify novel directions through extensive experimentation and refinement, a process that is both time- and cost-intensive.
Bringing a single new drug to market, for example, typically requires around 14 years and nearly \$2 billion~\cite{wouters2020estimated, hinkson2020accelerating}. 

A key bottleneck in this process is hypothesis generation, which requires researchers to reason across high-dimensional search spaces while balancing complex trade-offs among competing criteria~\cite{schunn19954}.
For example, in drug discovery, medicinal chemists must navigate a scaffold-constrained chemical space comprising approximately \(10^{4}\)--\(10^{6}\) possible analog molecules~\cite{nippa2025expediting, zin2018cheminformatics}, while balancing trade-offs among pharmacological properties such as absorption (e.g., PLogP~\cite{sun2023development}), toxicity (e.g., DILI~\cite{chen2015drug}), and blood--brain barrier permeability (e.g., BBBP~\cite{sun2023drug}).
Similarly, in materials science, researchers must explore approximately \(10^{10}\) possible quaternary compounds~\cite{davies2016computational} to optimize battery electrolytes, balancing ionic conductivity against electrochemical stability~\cite{sendek2017holistic}.
Although computational tools broaden exploration in these search spaces, they still offer only partial support for hypothesis formulation, iterative refinement, and reasoning about property trade-offs, leaving manual work to experts.

Recent advances in AI have opened new avenues for hypothesis generation.
Generative models now retrieve~\cite{li2025grappi, lin2025scirgen, qi2024large} and produce hypotheses~\cite{o2025sparks,cohrs2025large, zhou2024hypothesis} across diverse scientific tasks, including multi-property molecule optimization~\cite{dey2025gellm3o}, 3D protein-ligand docking~\cite{jumper2021highly}, and phenotypic profiling~\cite{chen2026harnessing}.
Interactive AI systems in HCI extend these capabilities to support research question formulation~\cite{pu2025ideasynth, liu2025personaflow}, related work retrieval~\cite{koloveas2025accelerating, lee2024paperweaver, zheng2024disciplink, portenoy2022bursting, kang2023comlittee}, and data analysis~\cite{khanal2024fathomgpt, shi2022medchemlens, gao2024collabcoder}, enabling researchers to engage with previously unexplored information.

However, to the best of our knowledge, no AI tools interactively support the reasoning process during hypothesis generation~\cite{pool2024ai}.
The lack of such support implies that researchers must manually and independently construct hypotheses through iterative reasoning over AI-provided information, indicating that the fundamental bottleneck of hypothesis generation still persists. Consequently, researchers may confine their search to familiar patterns, often overlooking novel and promising directions~\cite{foster2015tradition, simon2019sciences, amershi2019guidelines}.
Designing AI to actively support this reasoning process remains an open challenge in HCI.

In this paper, we present a co-abduction framework that extends abductive reasoning~\cite{peirce1934collected} to human-AI collaboration in hypothesis generation.
Researchers and AI jointly generate and cluster hypothesis candidates by property improvement profiles, identify strategies to improve properties within each cluster, and synthesize strategies across clusters to produce new and promising hypotheses.
We demonstrate our approach through ligand optimization in drug discovery\footnote{See {\S}\ref{bg} for more background information}, where researchers explore and develop molecular hypotheses while managing complex trade-offs among multiple pharmacological properties, such as improving solubility while preventing increased toxicity.

We conducted a formative study with five medicinal chemists to understand how they use AI in their reasoning.
Chemists use generative models (e.g., GeLLM$^{4}$O-C~\cite{dey2025large} and DrugGen~\cite{sheikholeslami2025druggen}) to produce improved molecules, group them by the improved property, and identify recurring fragment–property relationships.
They then recombine fragments across groups to generate hypotheses that balance multiple properties, connecting structural observations to broader explanatory patterns.
Our findings revealed that this workflow aligns with abductive reasoning~\cite{veen2021creative, peirce1934collected} across three stages: observation, pattern identification, and hypothesis generation.
However, each stage exposes gaps in current tool support: (1) filtering improved molecules lacks streamlined support, (2) identifying frequent fragments is labor-intensive while LLMs remain unreliable due to hallucination, and (3) LLMs lose coherence during iterative multi-fragment refinement.

Based on these insights, we developed \system (Figure~\ref{fig:first}), an interactive interface for human-AI co-abduction:
\begin{itemize}
    \item \textbf{Improved-candidate Cluster}: The system generates molecules that improve target properties and clusters them by their property-improvement profiles.
    \item \textbf{Strategy Identification}: Researchers explore clusters, and AI explains patterns and suggests strategies to improve worsened properties (intra-cluster strategies).
    \item \textbf{Multi-strategy Synthesis}: AI suggests strategies across clusters (inter-cluster strategies), and researchers iteratively recombine and refine them until the target properties are met.
\end{itemize}

We conducted an expert study with 10 medicinal chemists, comparing \system against a baseline with the same AI model but no abductive reasoning support.
Participants rated \system significantly higher in observation ($p$ < 0.01), strategy identification ($p$ < 0.01), multi-strategy composition ($p$ < 0.01), and overall satisfaction ($p$ < 0.01), as well as ligand optimization quality and diversity.
Usage analysis further revealed that participants frequently experienced abductive leaps (``aha!’’ moments~\cite{klahr1988dual,dunbar1995scientists}), after which they showed greater initiative and sharply reduced their reliance on AI in favor of manual editing.
These findings suggest that human-AI collaboration should adapt to the user’s cognitive stage: before an abductive leap, AI should act as a catalyst, providing divergent triggers; after the leap, AI should shift to a supporting role as users refine their realizations.

In summary, our key contributions are as follows:
\begin{itemize}
    \item \textit{Co-abduction}, a human-AI collaborative framework for abductive reasoning in scientific hypothesis generation, enabling efficient observation, systematic strategy identification, and coherent multi-strategy synthesis.
    \item Design and implementation of \system, an AI system that supports co-abduction for molecular hypothesis generation in ligand optimization.
    \item An expert study with 10 professionals demonstrating that iterative co-abduction significantly improves hypothesis quality and diversity.
    \item Design implications for building AI systems to support hypothesis generation in scientific domains.
\end{itemize}

\section{Related work}

\subsection{Hypothesis Generation in Scientific Domains}
Hypothesis generation is an iterative reasoning process central to scientific discovery~\cite{schickore2014scientific}, in which researchers explore data representations to develop and refine plausible explanations~\cite{klahr1988dual, schunn19954}.
Across scientific domains, this process requires navigating high-dimensional search spaces while managing trade-offs among competing criteria.
In drug discovery, the space of drug-like molecules spans $10^{23}$--$10^{60}$~\cite{polishchuk2013estimation, bilodeau2022generative}, with competing pharmacological properties such as stability, solubility, and toxicity~\cite{sendek2017holistic}.
Materials discovery confronts over $10^{12}$ inorganic compositions~\cite{davies2016computational}, requiring balance among conductivity, mechanical robustness, thermodynamic stability, synthesizability, and cost~\cite{jain2013commentary, butler2018machine}.
Bioscience faces still larger spaces, where a 100-residue protein sequence yields $20^{100} \approx 10^{130}$ possibilities~\cite{dryden2008much}, with fundamental trade-offs between bioactivity and stability~\cite{bloom2006protein,tokuriki2009stability}.

To navigate vast hypothesis spaces across multiple criteria, scientific communities have developed computational interfaces that integrate large-scale databases to support the retrieval and screening of validated findings. In drug discovery, platforms such as PubChem~\cite{pubchem2021} and ChEMBL~\cite{gaulton2012chembl} provide searchable access to chemical structures and bioactivity data. In materials science, resources like the MaterialsAtlas~\cite{materialsAtlas2022} and AFLOW~\cite{curtarolo2012aflow} enable filtering of inorganic compounds by stability and functional properties. In bioscience, repositories such as UniProt~\cite{uniprot2023} and the Protein Data Bank (PDB)~\cite{berman2000pdb} offer curated sequence and structural data for comparative analysis. Despite these advancements, most systems focus on retrieving and ranking previously validated data rather than supporting interactive reasoning, which is a core part of hypothesis generation.

\subsection{AI Systems for Hypothesis Generation}
Recent advances in AI models have begun to accelerate the hypothesis-generation process. Generative models show great potential for hypothesis generation by retrieving~\cite{li2025grappi, lin2025scirgen, qi2024large} and generating hypotheses~\cite{o2025sparks, cohrs2025large, zhou2024hypothesis}. These models enable systematic exploration of vast hypothesis spaces by synthesizing multimodal scientific data and identifying latent patterns that inform experimental design.
Building on these technical capabilities, specialized platforms are emerging that integrate diverse models to support the entire pipeline, from data preprocessing to model deployment.
For example, Nvidia BioNeMo~\cite{bionemo2024} is an AI platform for drug discovery that simplifies and accelerates the development and training of various AI models for drug discovery applications, such as AlphaFold~\cite{jumper2021highly}, DiffDock~\cite{corso2022diffdock}, and MegaMolBART~\cite{irwin2022chemformer}.

Building on these advances, interactive AI systems in HCI increasingly support hypothesis generation, including research question formulation, paper retrieval, and data analysis. For research question formulation, IdeaSynth enables iterative refinement of research briefs, allowing users to explore variations of initial ideas~\cite{pu2025ideasynth}, while PersonaFlow supports research question refinement through feedback from multiple domain expert perspectives~\cite{liu2025personaflow}.
For paper retrieval, BIP! Finder accelerates literature review through AI-assisted document synthesis~\cite{koloveas2025accelerating}, and ComLittee assembles expert committees to recommend papers based on members’ expertise~\cite{kang2023comlittee}. For data analysis, MedChemLens provides visual tools for extracting drug-related insights from existing records~\cite{shi2022medchemlens}, and FathomGPT supports interactive exploration and analysis of ocean science data, images, and figures~\cite{khanal2024fathomgpt}.

Across these tasks, such systems fundamentally support sensemaking: organizing and interpreting fragmented information into coherent representations that guide exploration~\cite{pirolli2005sensemaking}. 
Early collaborative systems supported sensemaking by helping people organize large sets of artifacts into shared structures such as taxonomies and conference schedules~\cite{chilton2013cascade, chilton2014frenzy, kim2013cobi}. More recently, interactive tools have extended sensemaking to AI-generated outputs, helping users compare, cluster, and evaluate large volumes of model responses~\cite{gero2024supporting, arawjo2024chainforge, glassman2015overcode, almeda2024prompting, shankar2024validates, gu2025abstractexplorer}, and structuring GenAI-assisted hypothesis exploration through shared visual representations~\cite{ding2025diagram}.

Although existing systems support sensemaking by broadening the search space, including finding, surfacing, and organizing candidates, they provide little support for reasoning about the trade-offs that determine which hypotheses are scientifically viable. As a result, researchers must still manually synthesize fragmented AI outputs to judge promising directions on their own. We introduce \system, an interactive interface for human-AI co-abduction in which researchers and AI collaboratively uncover explanations behind emerging patterns and apply them to generate well-grounded hypotheses.

\subsection{Co-abductive Reasoning: Synergizing Human and AI }
Abductive reasoning is the process of inferring the most plausible explanation for an observed phenomenon, whether diagnosing the cause of a failure or decoding the principles of success for further optimization~\cite{peirce1934collected}. 
Abduction functions not only as a logical form but also as a strategic methodology that integrates all available constraints and background knowledge into a coherent, organized pattern~\cite{kroll2017studying,paavola2004abduction}.
While deduction guarantees certainty and induction establishes probability~\cite{read1914logic, hitchcock1979deductive, shanahan2022deductive}, abduction provides the creative leap necessary to introduce novel hypotheses in the face of uncertainty~\cite{veen2021creative, peirce1934collected}, and stands as the only logical operation capable of generating truly new ideas~\cite{peirce1934collected, barrena2019abduction}.

The ultimate goal of abductive reasoning is the abductive leap, an insightful reconfiguration of observations into a plausible explanatory hypothesis~\cite{garbuio2021innovative}, often experienced as an Aha! moment~\cite{tik2018ultra, carpenter2019aha} of sudden understanding. This abductive leap does not result from creating entirely new hypotheses, but from strategically combining previously unconnected elements within a specific problem context and recognizing their validity anew~\cite{paavola2004abduction}. Conceptually, abductive reasoning generally unfolds in three stages~\cite{thagard2010brains, magnani2009creative}: 1) observing phenomena, 2) identifying patterns, and 3) recombining disjoint or overlapping patterns into novel hypotheses. For example, an observed improvement under one condition and a successful strategy from another context may be recombined into a new hypothesis explaining or extending both observations.

To gain a deeper understanding of such strategic abduction in the real world, we conducted a formative study with ten medicinal chemists. Building on these findings, we present co-abductive reasoning, in which humans and AI share a continuous abductive loop through three stages: (1) Improved-candidate Clustering, clustering hypothesis candidates by property improvement profiles; (2) Strategy Identification, distilling optimization strategies within a cluster; and (3) Multi-strategy Synthesis, combining strategies across clusters to iteratively refine molecules toward target properties.

\section{Background \& Definitions of Key Terminology}\label{bg}
In this section, we provide background on lead optimization and define key terminology to aid understanding.

\begin{figure}[H]
    \centering
    \includegraphics[width=1\linewidth]{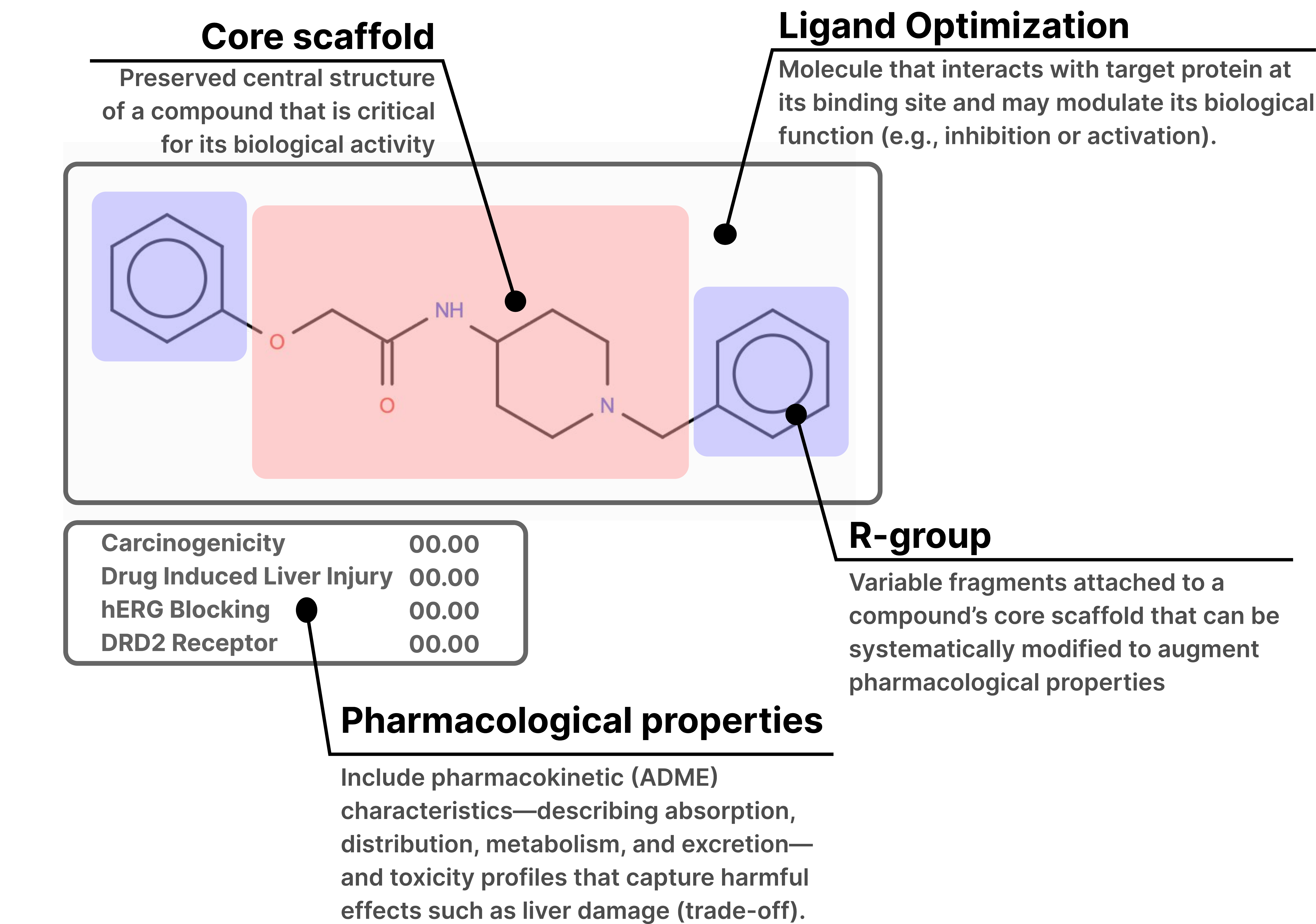}
    \caption{Description of key terminologies for compound optimization, including lead compound, pharmacological properties, core scaffold, and r-group.}
    \label{fig:definitions}
\end{figure}

\textbf{Ligand Optimization.}  Medicinal chemists generate molecular candidates (hypotheses) for wet-lab testing by designing an initial molecule that simultaneously improves multiple pharmacological properties with trade-off, including absorption (e.g., PlogP~\cite{sun2023development}), toxicity (e.g., LIVER~\cite{chen2015drug}), and blood–brain barrier permeability (e.g., BBBP~\cite{sun2023drug}). 

\textbf{Ligand.} A small molecule that interacts with a target protein at its binding site and may modulate its biological function (e.g., inhibition or activation)~\cite{kitchen2004docking}. 
Through high-throughput screening (HTS)—a rapid experimental method for testing large numbers of compounds-medicinal chemists identify those with high binding affinity to a target protein and desired biological activity~\cite{macarron2011impact}.

\textbf{Pharmacological Properties.} Pharmacological properties include pharmacokinetic (ADME) characteristics—describing absorption, distribution, metabolism, and excretion—and toxicity profiles that capture harmful effects such as liver damage (trade-off). For example, increasing solubility—the extent to which a drug dissolves in water or biological fluids—may simultaneously raise liver or cardiac toxicity, leading to adverse side effects~\cite{mceuen2017associations, chen2013high}.

\textbf{Core Scaffold.} Core scaffold refers to the preserved central structure of a compound that is critical for its biological activity. 
Therefore, medicinal researchers maintain the core while generating new compounds~\cite{schuffenhauer2007scaffold}.

\textbf{R-Group.} R-groups are the variable fragments attached to a compound’s core scaffold that can be systematically modified to augment pharmacological properties~\cite{maggiora2014molecular}.

\section{Formative study}\label{fs}
The formative study aimed to understand 
(1) how medicinal chemists perform ligand optimization in practice; 
(2) how AI tools are currently used during ligand optimization; 
and (3) the limitations of existing AI systems in supporting this task, which informed our design goals.

\subsection{Participant}
We recruited five participants with backgrounds in medicinal chemistry for the formative study: four senior researchers and one Ph.D. candidate, all currently working on ligand optimization (see Table~\ref{tab:participant} in the appendix). Their work experience ranged from 4 to 18 years ($M = 12.2$, $SD = 5.4$). Participants were recruited through personal referrals and by contacting research labs via email to request collaboration. The study was approved by the Institutional Review Board (IRB), and participants’ consent was obtained before the study. Each participant received a \$35 gift certificate. When reporting interview quotes, we use $P{n}$ to denote the participant number.

\subsection{Procedure}
The study consisted of three parts. First, we conducted a 30-minute in-depth interview covering (1) participants’ workflow in ligand optimization, (2) their goals and strategies when using AI tools, and (3) their past experiences with AI tools in ligand optimization, including challenges encountered. Next, a 20-minute workflow observation session provided deeper insight into their design strategies as participants used their preferred AI tools to optimize molecules they were currently working on. Finally, a 20-minute post-task interview focused on their refinement strategies and difficulties. Each session lasted 1 to 1.5 hours.
After completing the interviews, we applied thematic analysis and iterative open coding~\cite{clarke2015thematic} to analyze the interview transcripts. Three authors coded and analyzed the transcripts for emerging themes, and the findings were discussed among the co-authors iteratively until consensus was reached.

\subsection{Findings}\label{finding}
We identified that, to optimize molecules, researchers typically follow three stages ({\S}\ref{procedure}): (1) grouping molecules that exhibit property improvements, (2) extracting patterns between augmented properties and structural fragments, and (3) combining these patterns to design new candidate molecules with improved multi-property profiles. We found that this process aligns with abductive reasoning, a representative theory of hypothesis generation that moves from observation to pattern discovery to hypothesis formulation.

Building on this perspective, we examined both the role of AI ({\S}\ref{AIrole}) and its limitations within each abductive step ({\S}\ref{AIlimit}) and, in turn, derived our system design goals to better support hypothesis-driven molecular optimization ({\S}\ref{DG}).

\subsubsection{Molecular Hypothesis generation in lead optimization}\label{procedure} 
We identified three stages in ligand optimization: (1) grouping molecules that show property improvements, (2) extracting patterns between augmented properties and structural fragments, and (3) combining these patterns to design new candidate molecules with improved multi-property profiles.

First, medicinal chemists identify candidates for the R-group and group them based on their impact on specific pharmacological properties. They determine the core scaffold of an initial lead compound that must be preserved, as well as the R-groups that can be modified. 
They then search for prior optimization cases with the target properties. 
Researchers often identify molecules in which specific target properties have been notably improved, such as one or two out of four, and group them accordingly. Next, they determine which fragments contributed to improvements in particular properties. Within each group of molecules, fragments that appear most frequently are considered more likely to be associated with the observed property improvements. For example, molecules that score highly in solubility and toxicity mitigation often contain specific polar functional groups, whereas molecules with improved permeability and potency frequently include hydrophobic or aromatic substituents~\cite{cisneros2017systematic, he2024innovative, alkharboush2025fragment}. Such relationships between specific fragments and property improvements provide valuable strategies for researchers aiming to optimize multiple properties simultaneously. Based on these findings, they design and evaluate compounds through R-group optimization~\cite{blower2004systematic}, combining the preserved core scaffold with selected fragments at modifiable positions (R-groups). They define several positions (R\textsubscript{1}, R\textsubscript{2}, R\textsubscript{3}) on the scaffold and systematically assign the previously identified fragments to these sites, generating a series of analogs with different property profiles. 

This process, in a small-scale synthesis setting, typically takes at least one week. In the first round, researchers design approximately 7 to 10 candidate molecules and evaluate their properties through wet-lab experiments. In the second round, based on these results, they iteratively refine and expand promising candidates, ultimately generating a larger set of over 20 molecules for further exploration.

\subsubsection{Roles of AI in Ligand Optimization}\label{AIrole} 
Medicinal chemists often use specialized AI models to design and evaluate new molecules. In addition, large language models have been applied to reasoning strategies for optimization, such as grouping improved molecules, distilling strategies from these groups, and integrating the strategies into a final design.

All participants are familiar with using molecular generation models, such as GeLLM$^{4}$O-C~\cite{dey2025large} and DrugGen~\cite{sheikholeslami2025druggen}, to directly explore novel chemical space. These models generate new molecules to simultaneously optimize multiple properties, given the initial molecule and the targeted properties. Researchers efficiently explore vast molecular spaces using generative AI models.
For example, \hl{\textit{“Generative models can easily generate new molecules to satisfy properties. Although many of them are difficult to synthesize, they often provide ideas or structural variations that we would not have considered.”}} ($P{2}$). However, these tools often require programming knowledge and familiarity with computational workflows, creating a practical barrier for many experimental chemists. \hl{\textit{“Many generative chemistry models have great potential, but they are hard to use in practice for researchers who do not have coding experience”}} ($P{4}$).
\hl{\textit{“...ChatGPT and Gemini might have potential in generating molecules; for now, the results are not that promising.”}} ($P{3}$).

For evaluation, the overall pharmaceutical property scores of these generated molecules are evaluated with AI-based interfaces such as SwissADME~\footnote{http://www.swissadme.ch} and ADMET LAB~\footnote{https://admetlab3.scbdd.com/}, with real-time evaluation. Medicinal chemists manually edit the generated molecules and repeat the process until the desired properties are achieved.

With recent advances in LLM capabilities, many researchers expect these systems to move beyond generation and evaluation to support reasoning for new molecular hypotheses. In practice, they frequently consult LLM interfaces such as ChatGPT\footnote{\url{https://chatgpt.com/}} and Gemini\footnote{\url{https://gemini.google.com/}}. Researchers often input multiple improved molecules into the model and ask it to group and distill the most frequent fragment patterns. \hl{\textit{“Once we fix the scaffold, I often ask an LLM to suggest plausible R-group substitutions that could modulate potency, selectivity, or ADMET properties to understand some patterns between augmentation and molecular structures.”}} ($P_1$) To combine multiple strategies, researchers often input pairs of group-augmented molecules into an LLM to explore potential combinations of modifications. Although hallucination remains a concern, the model occasionally suggests meaningful and novel combinations. Despite this potential, LLM outputs still require substantial manual refinement. For example, \hl{\textit{“The model can propose interesting combinations, but we still need to manually evaluate whether they make chemical sense and whether the properties actually improve.”}} ($P_3$)

\subsubsection{AI Optimization Workflows and Key Challenges}\label{AIlimit} 

In the workflow observation, their workflows were similar: they generate multiple molecules, identify relationships between augmenting properties and specific fragments, and iteratively refine the initial molecule by integrating these relationships.

They usually begin by grouping molecules whose properties have been improved (improved molecules).
However, a key challenge was the lack of a streamlined way to filter for improved molecules; consequently, researchers had to manually summarize the AI output from generation and evaluation for multiple molecules.
\hl{\textit{``When I generate hundreds of variants from fragment substitutions, I have to evaluate each one individually and manually prioritize them, which makes the process overwhelming''}}($P{3}$). 
This labor-intensive process prevents them from fully leveraging generation and evaluation models. These challenges highlight the inefficiencies caused by the disconnect between molecular generation and the discovery of meaningful patterns.

Another critical bottleneck is that standard LLMs lack spatial grounding; they cannot visually identify which specific regions of a molecule are functionally significant. Because LLMs primarily operate on text-based representations such as SMILES~\cite{weininger1988smiles}, they cannot intuitively demonstrate how a fragment’s efficacy varies with its precise spatial positioning.
\hl{\textit{“Even if we have a list of molecules, we still need to manually map each fragment to R$_1$, R$_2$, or R$_3$ before we can analyze which fragments are actually driving the property changes.”}} ($P{4}$).
Consequently, researchers must manually inspect molecular structures to determine where specific fragments can be applied. Given that a fragment's impact can change drastically depending on its location, this limitation prevents experts from effectively leveraging LLMs to clearly identify patterns between a fragment and a property.

The final major challenge is maintaining context when integrating multiple strategies. Even with grouped molecules and extracted patterns, LLMs often lose track of these complex constraints during the iterative refinement of multi-pattern combinations.
"\hl{\textit{“When I try to combine a strategy for improving solubility with another for enhancing potency, the model often 'forgets' the original scaffold constraints or swaps functional groups in ways that make no sense. It’s like it loses the big picture while focusing on a specific edit.”}} ($P_{2}$).
Consequently, this lack of context retention can result in chemically implausible structures or hallucinated suggestions that violate fundamental chemical principles or predefined optimization goals.

\begin{figure*}
\centering
  \includegraphics[width=1.9\columnwidth]{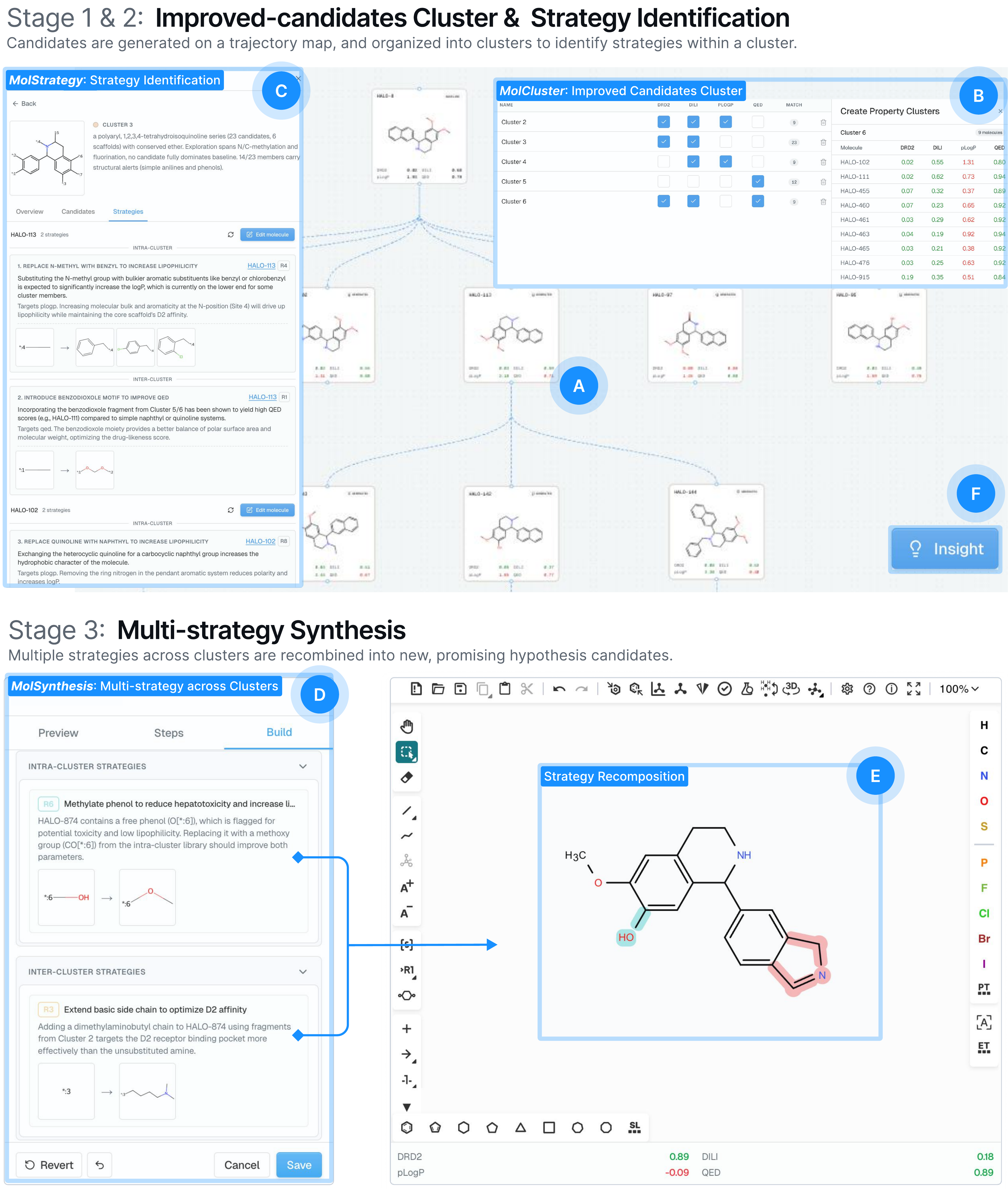}
  \caption{Overview of the \system workflow: (1) Improved Candidates Cluster, (2) Strategy Identification, and (3) Multi-strategy Synthesis. Improved Candidates Cluster (A, B) generated molecules and organized them into property-based clusters to support efficient observation. Strategy Identification (C) extracts and summarizes intra-cluster optimization strategies from candidate groups. Multi-strategy Synthesis (D, E) enables users to recombine multiple intra-and inter-strategies into actionable edits to generate new, promising candidates. The Insight button (F) allows users to externalize and reflect on abductive leaps as they occur throughout the workflow.}

  ~\label{fig:HALO}
\end{figure*}

\subsection{Design goals}\label{DG} 
Based on these findings, we proposed three design goals for our system:
\begin{itemize}

\item \textbf{DG1: Efficient Observation through Grouping of Improved Molecules.} The system should enable researchers to group improved molecules by diverse property combinations, intuitively connecting molecular generation with the discovery of meaningful patterns without labor-intensive manual inspection.

\item \textbf{DG2: Systematic Strategy Identification through Pattern Formalization within Groups.}
The system should provide explicit guidance on understanding patterns between a fragment and a property, specifically which fragments are used for modification and where to position them for R-group design, thereby reducing reliance on manual and heuristic inspection of molecules. 

\item \textbf{DG3: Coherent Multi-Strategy Synthesis across Groups.} The system should maintain structural and contextual coherence as researchers recombine and refine multiple strategies across groups by simplifying the assembly of complex molecule combinations and reducing the cognitive load of merging complex optimization paths.

\end{itemize}

\section{\system: Human-AI Co-abduction system for Lead Optimization}

Building on these design goals (DG1-3), we present \system, a human-AI co-abduction system for lead optimization through three main components: 1) \fone, 2) \ftwo, and 3) \fthree. We explain the technical details of three components in {\S}\ref{techdetails} in the appendix.

\subsection{\fone: Improved Candidates Cluster}\label{sec:stage1}

\fone (Figure~\ref{fig:HALO}-B) is designed to support efficient observation by enabling users to explore which properties have improved or worsened during AI-driven generation, forming an initial understanding of the optimization landscape.
Users can explore and organize diverse molecule clusters as a foundation for subsequent strategy identification and synthesis.
The workflow proceeds as follows:
\begin{itemize}[leftmargin=0.25in]
\item Step 1-1: Users first generate dozens of candidate molecules from an initial molecule on an interactive trajectory map (Figure~\ref{fig:HALO}-A) using a generative AI model or by manual editing. The trajectory map allows them to observe how molecular structures evolve and how these changes affect the property score.
\item Step 1-2: The \fone (Figure~\ref{fig:HALO}-B) enables multi-criteria screening of the candidate set, grouping molecules into clusters based on combinations of improved properties (\textcolor{improved_green}{green: augmented} and \textcolor{worsened_red}{red: worsened}).
\item Step 1-3: By creating these clusters, users develop an understanding of the landscape of generated molecules. They identify worsened properties and use these insights to guide subsequent optimization steps.
\end{itemize}

\subsection{\ftwo: Systematic Strategy Identification}\label{sec:stage2}
\ftwo (Figure~\ref{fig:HALO}-C) is designed to help users systematically identify optimization strategies within a cluster by exposing detailed cluster-level insights.
The workflow proceeds as follows:
\begin{itemize}[leftmargin=0.25in]
\item Step 2-1: The overview tab (Figure~\ref{fig:mol_sys} in the appendix) displays an AI-generated summary of the cluster alongside its identified strengths and weaknesses to help users understand an cluster overall.
\item Step 2-2: The candidates tab (Figure~\ref{fig:mol_sys} in the appendix) displays all cluster members in a sortable table with absolute property values and improvement deltas relative to the initial molecule.
\item Step 2-3: The strategies tab summarizes actionable strategies for improving worsened properties within the cluster, suggesting which molecular fragments of cluster members should be modified to address observed weaknesses. If users click a molecule in the list, they can open \fthree to edit it (Figure~\ref{fig:HALO}-C).

\end{itemize}

\subsection{\fthree: Coherent Multi-strategy Synthesis}\label{sec:stage3}
\fthree (Figure~\ref{fig:HALO}-D and E) is designed to help users synthesize multiple strategies across clusters. 
The workflow proceeds as follows:
\begin{itemize}[leftmargin=0.25in]
\item Step 3-1: \fthree provides users with both intra-cluster and inter-cluster strategies with specific explanations, enabling users to explore diverse fragments for optimization (Figure~\ref{fig:HALO}-D). 
\item Step 3-2: When users click on a particular strategy, the molecule on the molecule editor (Figure~\ref{fig:HALO}-E) is updated to reflect the strategies.
\item Step 3-3: Users can check the change of property scores at the bottom (\textcolor{improved_green}{green: augmented} and \textcolor{worsened_red}{red: worsened}), and with the change, they iteratively refine the molecules until more properties are satisfied.
\item Step 3-4: Saving creates a new node on the tree, connected to its source by an edit edge, which can serve as the starting point for a new round of generation (Figure~\ref{fig:HALO}-A).

\end{itemize}

We expect that \system supports co-abduction through three stages: clustering improved candidates, distilling optimization strategies within a cluster, and synthesizing multiple strategies across clusters, thereby helping users generate new, promising molecular hypotheses.

\section{User Study}\label{US}
Our user study evaluates how \system supports professionals in abductive reasoning within their workflows: 1) grouping improved molecules, 2) extracting strategies from groups, and 3) merging strategies into a new one. We conducted a within-subjects experiment simulating real-world ligand-optimization workflows.

To isolate and evaluate co-abductive reasoning, we developed a baseline system ({\S}\ref{baselinedesign} in the appendix) that includes the non-co-abductive features of \system (e.g., trajectory map, AI/manual molecule generator, molecular information, and LLM search) but excludes three co-abductive features (\fone, \ftwo, and \fthree). 
We designed the baseline based on insights from the formative study ({\S}\ref{finding}) that show AI-assisted practice, in which experts use generative models and free-form prompting to generate, evaluate, and reason over candidates. This controlled comparison allows us to assess the contribution of co-abduction in \system to molecular hypothesis-generation workflows.

Our study aimed to answer the following research questions:

\begin{itemize}

\item \textbf{RQ1-1}: Can \system help researchers efficiently identify which properties are improved or worsened by grouping improved molecules?

\item \textbf{RQ1-2}: Can \system support researchers in systematically identifying optimization strategies within groups of improved molecules?

\item \textbf{RQ1-3}: Can \system enable researchers to coherently recombine multiple optimization strategies to generate new candidate molecules?

\item \textbf{RQ2}: 
To what extent does \system achieve multi-objective property enhancement while maintaining structural diversity compared to baselines?

\item \textbf{RQ3}: 
Does an abductive leap occur during the use of \system, and if so, how does it transform human–AI interaction in the discovery trajectory?

\end{itemize}

\subsection{Study Design}

\subsubsection{Task}\label{tasks} 
We designed two tasks, Task A and Task B (Figure~\ref{fig:task_ex} in the appendix), in which participants improve the initial molecule with respect to four properties ({\S}\ref{property} in the appendix).
The tasks were based on two criteria: 1) versatility of the initial lead compound and 2) applicability of the properties to real-world experimental contexts.

First, in selecting the initial lead compound, versatility was essential, as the study does not focus on a specific drug or disease. To ensure diverse modification attempts, we chose the most frequent molecules from a dataset of 100,000 commercially available molecules, where property variations were generated through fragment modification~\cite{chen2021deep}. Each data instance contains a before-molecule with its property scores and an after-molecule with updated property scores generated by fragment modification. Their frequent occurrence indicates strong potential for diverse applications in lead optimization. We selected the top 10 instances from the dataset.

Second, for the desired properties, we emphasized relevance to real-world experimental contexts, where multiple properties often involve trade-offs that make simultaneous optimization challenging. To reflect this complexity, we selected properties with well-established trade-off relationships, specifically between PlogP and liver toxicity. Higher PlogP values are often associated with increased metabolic load and an elevated risk of liver injury~\cite{mceuen2017associations, chen2013high}, thereby reflecting the difficulty of optimization in practice. 
We selected two cases from the top 10 instances. By grounding the task in the most frequent lead compound and representative properties, participants explore a subspace smaller than the molecule hypothesis space required in real-world lead optimization tasks~\cite{nippa2025expediting, zin2018cheminformatics} (\(\sim\!10^{4}\)--\(10^{6}\) ).

\subsubsection{Procedure}
Participants were asked to complete both the experimental and control tasks. To control for potential order effects, we counterbalanced the sequence of conditions so that half of the participants began with the experimental condition and the other half with the baseline. We also counterbalanced the assignment of Tasks A and B across conditions so that each task was performed equally often under both the experimental and control settings. Each task was limited to 30 minutes, with a five-minute break between tasks. Before starting each task, participants received a 10-minute instruction session on how to use both corresponding interfaces and were given an additional five minutes to practice using the system. The study was conducted remotely via Zoom\footnote{https://www.zoom.com/}, and participants were asked to share their screens during the timed tasks. As the final output, participants were asked to submit the top seven molecules in the first-round setting, building on the formative study ({\S}\ref{procedure}).

\subsubsection{Measurement}
\begin{figure}
\centering
  \includegraphics[width=1.05\columnwidth]{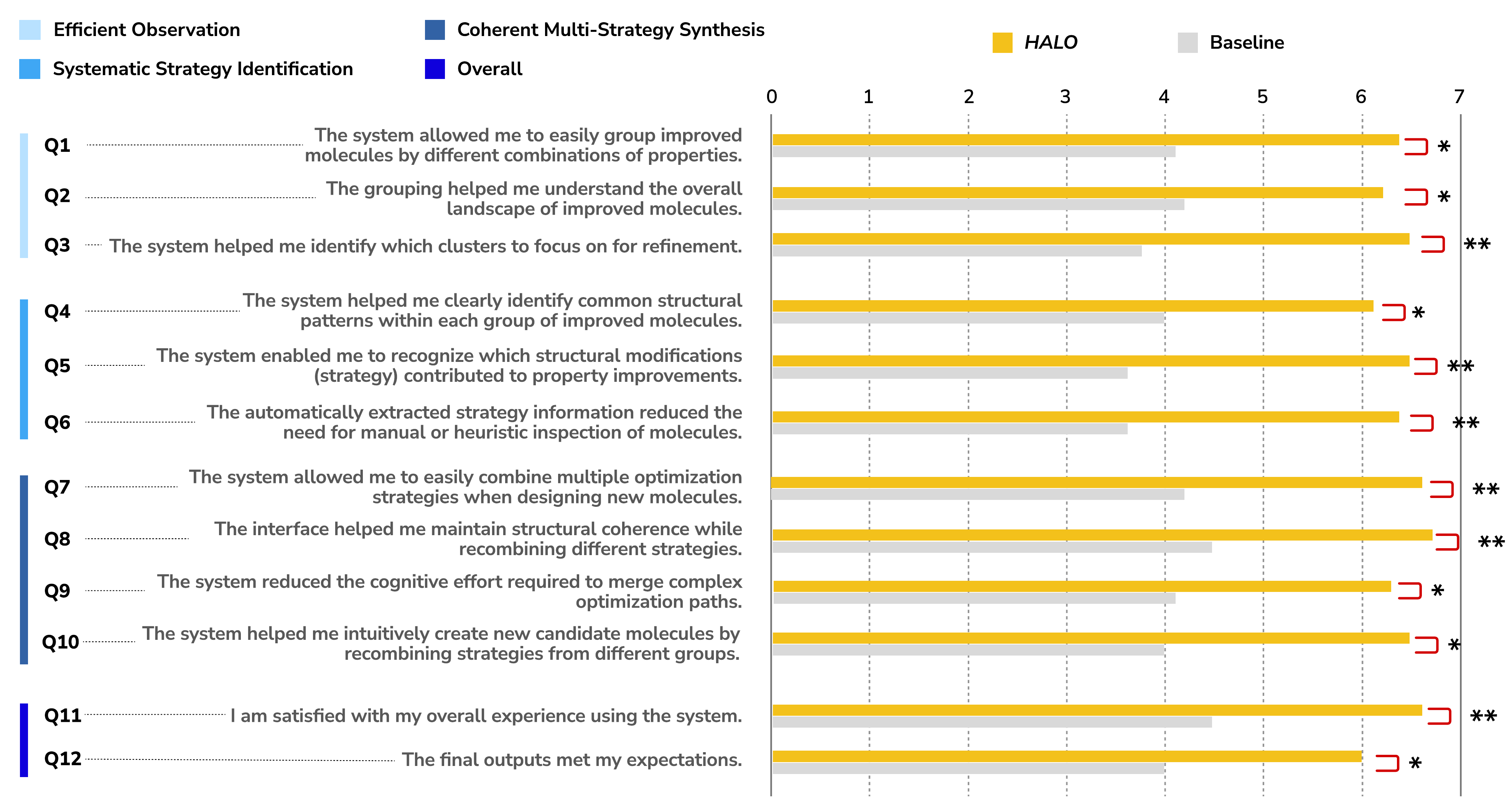}

  \caption{Bar plots showing the results of the user study. 
  Participants rated efficient observation (Q1-3), systematic strategy identification (Q4-6), and cogerent multi-strategy synthesis (Q7-10), and overall evaluation (Q11-12) for both the baseline and \system using a 7-point Likert scale ($^{*}p<0.05$, $^{**}p<0.01$, $^{***}p<0.001$). 
  }
  ~\label{fig:results}
\end{figure}

The post-condition questionnaire evaluated three research questions (R1-1, R1-2, R1-3): efficient observation, systematic strategy identification, coherent strategy composition, and overall user experience. 
These twelve questions were developed based on insights from the formative study ({\S}\ref{fs}). All questions and results are shown in Figure~\ref{fig:results}. 
Users rated each task in the survey on a 7-point Likert scale (1 is ``Not at all'' and 7 is ``Very much''). 
For RQ2, we evaluate the quality and diversity of submitted molecules: 1) Quality, measured by how many of the four properties were improved in each submitted compound; 2) Diversity, measured using pairwise structure similarity between submitted molecules. 
For RQ3, we analyzed log data capturing abductive leaps. 
Building on prior work~\cite {carpenter2019aha}, participants were instructed to click the Insight button (Figure~\ref{fig:HALO}-F) whenever they experienced an abductive leap during the experimental tasks with \system. 
We defined and explained an abductive leap as a moment when participants form a plausible, novel explanatory hypothesis or research direction.

\subsubsection{Participants}
We recruited ten professional medicinal chemists working in nine different drug discovery labs. Table~\ref{tab:participant} in the appendix presents the participants' demographic information. Their career experience ranged from 8 to 15 years ($M=10.90, SD=2.28$). We used snowball sampling, leveraging professionals who participated in the formative study. The Institutional Review Board (IRB) approved our study, and we obtained participants' consent before the study. Each participant received a \$75 gift certificate.

\subsection{Results and Findings}
We report findings organized by our research questions, combining results from the user study.

\subsubsection{\textbf{RQ1-1: Efficient Observation}}\label{RQ11new}

Participants using \system demonstrated more efficient observation rather than the baseline (Figure~\ref{fig:results}; Q1: $p < .05$,  Q2: $p < .05$,  Q3: $p < .01$).
Many participants ($P{5}$–$P{6}$, $P{9}$–$P{12}$) noted that \system facilitated the grouping of improved molecules, enabling faster understanding of the overall landscape and more effective identification of clusters for refinement
As $P{5}$ explained, \hl{\textit{``Clustering based on the target property was pretty efficient and reasonable, and it matched how I’d group things naturally.''}} $P{9}$ added, \hl{\textit{``It was easy to see that the AI-generated molecules tended to have low PlogP, which helped me understand which property to pay closer attention to in the next phase.''}}

\begin{table}[t]
\centering
\caption{Comparison of Quality and Diversity of submitted molecules between conditions. Values indicate mean $\pm$ standard deviation ($^{**}p<0.01$).}
\begin{tabular}{cccc}
\toprule
Metric & Baseline & HALO & $p$-value \\
\midrule
Quality & $2.25 \pm 0.48$ & $3.16 \pm 0.37$ & $0.0039^{**}$\\
Diversity & $58.27\% \pm 11.01\%$ & $46.07\% \pm 7.68\%$ & $0.0039^{**}$ \\
\bottomrule
\end{tabular}
\label{tab:score}
\end{table}

\subsubsection{\textbf{RQ1-2: Systematic Strategy Identification}}\label{RQ12new}
\system significantly enhanced the systematic identification of strategies by identifying relationship fragments and property improvement, significantly reducing the manual effort and improving diversity (Figure~\ref{fig:results}; Q4: $p < .05$,  Q5: $p < .01$,  Q6: $p < .01$).
Many participants ($P_{3}$–$P_{8}$, $P_{11}$) noted that \system effectively identified rational optimization strategies that aligned with expert chemical intuition, reducing manual labor required in traditional workflows. As $P_{3}$ explained, \hl{\textit{``When improving PlogP, it was easy to plan the optimization direction by examining fragments that frequently appeared in clusters of molecules with higher PlogP.''}} $P_{11}$ further noted \hl{\textit{``The strategies provided alongside existing knowledge were generally reasonable, so we don't waste time in the lab, and additional literature review.''}}   
Others found that suggesting strategies enables users to develop a molecular hypothesis in diverse ways. $P_{8}$ further noted \hl{\textit{``Instead of just going with intuition, having multiple evidence-based strategies helped me think of things I wouldn’t have considered and explore different directions.''}}

\subsubsection{\textbf{RQ1-3: Coherent Multi-Strategy Synthesis }}\label{RQ13new}
\system significantly enables participants to maintain coherence in synthesizing strategies across multiple clusters. 
\system streamlined the design of new candidates by enabling the intuitive recombination of optimization strategies, maintaining structural coherence during the process (Figure~\ref{fig:results}; Q7: $p < .01$,  Q8: $p < .01$,  Q9: $p < .05$, Q10: $p < .05$).
Several noted that the modular nature of \system provided a 'plug-and-play' experience, allowing participants to easily recombine various strategies while maintaining structural coherence ($P_{3}$–$P_{8}$, $P_{11}$–$P_{12}$).
As $P_{6}$ explained, \hl{\textit{``I really liked the ability to combine fragments like LEGO blocks with just a click, while seeing the score update in real time, which allowed me to quickly design a variety of new molecules.''}} $P{9}$ added, \hl{\textit{``With ChatGPT or Gemini, if I tried to modify multiple fragments, I would probably lose track of the flow pretty quickly. \system, on the other hand, helped me continuously build new combinations.''}}
However, several participants noted that once the optimization direction became more concrete, \system tended to present too many strategy suggestions, which could feel overwhelming and hinder further refinement ($P_{8}$-$P_{12}$).
As $P_{12}$ remarked, \hl{\textit{``Sometimes I just want to make some very simple modifications. This tool can make suggestions for a lot of types of analogs.''}}

Overall, these results for RQs (RQ1-1, RQ1-2, and RQ1-3) suggest that \system supports abductive reasoning in molecular hypothesis workflows, highlighting the suitability of \system for such processes.

\subsubsection{\textbf{RQ2: Ligand Optimization} }\label{RQ2new}
We begin the discussion of the results with one of the most critical questions: did \system result in not only higher-quality but also a greater diversity in ligand optimization?
For quality, we evaluated how many of the four properties were improved, using ADMET-AI~\cite{swanson2024admet}, which is also used in \system ($\Delta = \text{after} - \text{before}$).
We employed the Wilcoxon signed-rank test~\cite{wilcoxon1945individual}. Table~\ref{tab:score} shows that there is a significant difference in quality between conditions ($p = 0.0039$).  

For diversity, we computed the average pairwise similarity within each set (where each participant submitted seven molecules), and then compared between conditions using the Wilcoxon rank-sum test~\cite{wilcoxon1945individual}. Specifically, for a set of $N$ molecules, the diversity score is defined as:

\begin{equation}
Average Similarity = \frac{1}{N} \sum_{i=1}^{N} \left( \frac{1}{N} \sum_{j \ne i} \text{sim}(m_i, m_j) \right)
\end{equation}

where $m_i$ and $m_j$ denote molecules in the set, $\text{sim}(\cdot, \cdot)$ represents Tanimoto similarity~\cite{sokal1963principles} between two molecules, and $N$ is the number of molecules in the set (i.e., $N=7$ in our study).
Table~\ref{tab:score} shows that there is a significant difference in diversity between conditions ($p = 0.0039$).

Overall, these results suggest that \system supports users in achieving both higher-quality optimization and more diverse exploration, highlighting its potential to facilitate the generation of novel and promising molecular hypotheses.

\begin{figure}
    \centering
    \includegraphics[width=1\linewidth]{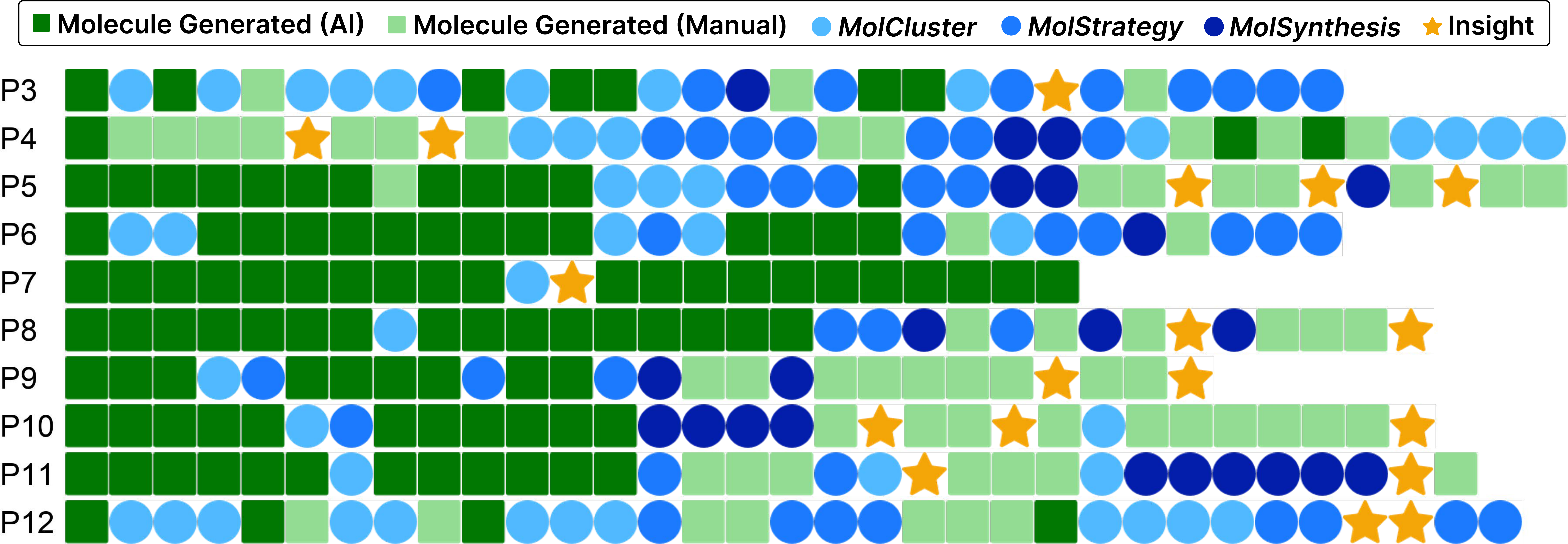}
    \caption{Each row illustrates a participant's component-usage logs recorded during the experimental task of the user study. Green squares represent the creation of new molecules, either through manual edits or molecular AI generations. Light blue circles represent \fone, blue circles represent \ftwo, and dark blue circles represent \fthree. The yellow star represents an Insight. Table \ref{tab:user_log_definition} in the appendix defines each label.}
    \label{fig:placeholder}
\end{figure}

\subsubsection{\textbf{RQ3: Abductive Leaps}}\label{RQ3new}

To answer RQ3, we analyzed the usage log of \system, including the Insight button (Figure~\ref{fig:HALO}-F), which participants clicked whenever they experienced an abductive leap toward a plausible and novel hypothesis or research direction.
Figure~\ref{fig:placeholder} shows participants' component usage log during the experimental task.
Participants reported, on average, 1.8 (SD = 0.92) abductive leaps (min: 0, max: 3) in usage of \system based on the log of the Insight button (Figure~\ref{fig:HALO}-F). 
We defined pre- and post-leap phases based on the first detected abductive leap and analyzed usage patterns in each phase.
In the pre-leap phase, abductive leaps often emerged when participants had experienced all three stages—efficient observation (\fone, {\fontsize{15}{15}\selectfont \textcolor{blue3}{$\bullet$}}, Figure~\ref{fig:placeholder}), strategy identification (\ftwo, {\fontsize{15}{15}\selectfont \textcolor{blue2}{$\bullet$}}, Figure~\ref{fig:placeholder}), and multi-strategy synthesis (\fthree, {\fontsize{15}{15}\selectfont \textcolor{blue1}{$\bullet$}}, Figure~\ref{fig:placeholder}). In particular, during the multi-strategy synthesis stage, participants often experienced a series of consecutive abductive leaps, suggesting an iterative refinement and expansion of emerging hypotheses rather than a single isolated insight.
For example, $P_{3}$ explained,
\hl{\textit{``I first had a new idea while trying to merge several strategies around cluster 1. Then, as I merged additional information around clusters 2 and 3, that original idea evolved further.''}} and $P_{6}$ noted, \hl{\textit{``
After diving into the whole molecule and the overall strategy, the logic behind the AI’s suggestions became clear. It actually inspired me to develop new approaches myself.''}} 
These findings suggest the importance of supporting abductive reasoning across the three-stage cycle in \system, while highlighting the organic interplay and interdependence among the stages.

In the post-leap phase, for generating molecules, AI usage (\colorbox{darkgreen}{\textcolor{darkgreen}{\rule{0.8em}{0.8em}}}, Figure~\ref{fig:placeholder}) dropped sharply (pre-leap: M=9.11, SD=4.54; post-leap: M=1.44, SD=3.68), while manual generation (\colorbox{lightgreen}{\textcolor{lightgreen}{\rule{0.8em}{0.8em}}}, Figure~\ref{fig:placeholder}) increased (pre-leap: M=3.33, SD=2.40; post-leap: M=4.11, SD=3.76)
For example, $P{5}$ explained, 
\hl{\textit{``Once the new idea came to mind, I wanted to refine it using my own knowledge,” one participant noted. “For an important task, if I triggered a new AI generation at that point, it felt like I would have to start over from scratch.''}} $P{7}$ remarked, 
\hl{\textit{``Because I cannot give the AI sufficiently detailed and precise instructions on my new idea, I preferred to take the lead and refine the idea myself after that new stage was reached.''}}
These findings suggest that participants primarily appropriated AI as a catalyst for initial conceptual leaps, but once a promising idea had emerged, they tended to reclaim control over the subsequent refinement process to preserve their intended direction.

\section{Discussion}
In this section, we reflect on our findings from the design and evaluation of \system. We also discuss design implications and future opportunities for improving and extending \system. 


\subsection{Design Implication 1: Generalizing \system-Co-abduction in Scientific Discovery}

We introduce a co-abduction framework that enables abductive reasoning in human–AI collaboration, incorporating three stages: (1) efficient observation, (2) strategy identification, and (3) multi-strategy synthesis. 
\system enables users to explore improved candidates in vast molecular spaces, organize them into meaningful groups, explain the reasons for their improvements, and identify, refine, and combine optimization strategies across these groups. 
Our three-stage framework improves quality and diversity of molecular optimization in drug discovery by supporting key processes of abductive reasoning, from fragmented observations to a coherent, unified hypothesis~\cite{alkan2025survey, zenil2026future}, presenting one possible operationalization of abductive reasoning in the real world.
Understanding the mechanisms behind optimization strategies and easily combining multiple strategies significantly enhances researchers’ sense of control, transparency, trust, and alignment with creative intent in scientific discovery, supporting core principles of human-AI interaction~\cite{wilkens2023configurations, schut2025bridging, shi2026survey, jiang2024human}.

Beyond drug discovery, \system’s implementation can be directly applied to other complex scientific domains that require abductive reasoning to generate promising hypotheses, including battery materials design and protein antibody design, where experts iteratively refine hypotheses by synthesizing diverse optimization strategies. 
For example, in battery materials design, material hypothesis (candidates) compositions can be clustered by improvement profiles across energy density, cycle life, and thermal stability (Stage 1: Improved-candidate Cluster); strategies can be identified within each cluster (e.g., adding dopants to offset stability loss) (Stage 2: Strategy Identification); and strategies are recombined across clusters to new materials (Stage 3: Multi-strategy Synthesis).

\subsection{Design Implication 2: Structuring AI's Imperfection into Grounded Reasoning}

Creative domains often see AI's imperfections as sources of novelty. AI often helps users break creative ruts by offering divergent options~\cite{martelaro2025inkspire, donahue2025amuse}; even structurally imperfect outputs can inspire new ideas. Jeon et al.~\cite{jeon2024weaving} reported that automotive designers applied twisted and blurred AI-generated designs as inspiration for new designs. 
In science, however, new discoveries demand objective grounding, such as satisfying multiple properties simultaneously. This means that such imperfections should be treated not as raw inspiration but as additional starting points for promising reasoning

Our study suggests how AI's imperfections can be structured into scientific hypothesis exploration through human-AI collaboration (human-in-the-loop~\cite{amershi2014power}). AI rarely produces a complete hypothesis that satisfies all properties in a single step; yet, when accumulated, these outputs form fragmented signals that point in a direction. 
Scientists further develop these signals into promising molecular hypotheses with their professional domain knowledge.
Clustering improved candidates yields a legible landscape for efficient observation (RQ1-1; $P_{5}$, $P_{9}$); surfacing fragment-level regularities turns weak associations into evidence-based directions beyond a user's intuition, expanding diversity (RQ1-2; $P_{3}$, $P_{8}$); and modular recombination with real-time feedback lets users confirm directions incrementally, sustaining coherence across synthesis (RQ1-3; $P_{6}$, $P_{9}$). In short, scientists achieve scientific leaps through structuring AI's incompleteness into grounded reasoning, improving both the quality and diversity of molecular hypotheses (RQ2).

This structuring of fragmented AI signals parallels sensemaking, organizing scattered information into coherent representations that guide exploration~\cite{pirolli2005sensemaking}. Given this overlap, co-abduction can draw on established sensemaking design patterns, from organizing artifacts into shared structures~\cite{chilton2013cascade, chilton2014frenzy, kim2013cobi} to comparing and evaluating large volumes of imperfect yet potentially valuable AI-generated outputs~\cite{gero2024supporting, arawjo2024chainforge, glassman2015overcode, almeda2024prompting, shankar2024validates, gu2025abstractexplorer, ding2025diagram}.

\subsection{Design Implication 3: Before and After Abductive Leaps}

The purpose of co-abduction is to facilitate an abductive leap in which a scientist suddenly reconfigures the problem space into a new, promising hypothesis~\cite{klahr1988dual,dunbar1995scientists}, such as an Aha! or eureka moment.~\cite{fedor2015problem,osuna2021current}.
Previous research has demonstrated that scientific approaches differ before and after the abductive leap~\cite{fedor2015problem,osuna2021current}.
Prior to the insight, scientists engage in iterative cycles of divergent thinking. They use technology to broadly explore the hypothesis space and iteratively build local conceptual contexts, continuing until scientists perceive sufficiently promising cognitive breakthroughs have emerged~\cite{klahr1988dual,dunbar1993concept,ohlsson1992information}.
Post-insight, the cognitive regime shifts: convergent thinking becomes dominant as scientists focus on refining and optimizing these ideas to meet predefined objectives~\cite{klahr1988dual,dunbar1993concept}.

In our results, while scientists often use AI to help form a promising explanatory frame beforehand, they rely on domain-specific sense-making rather than using AI afterward.
These differences in AI usage suggest a phase-dependent initiative in human–AI collaboration. Before insight, AI can have initiative comparable to that of humans, proactively supporting the iterative cycles of divergent and convergent thinking by understanding the user's needs and enabling progress toward a crystallized explanatory framework.
After insight, the initiative shifts back to the human, and AI should adopt a more restrained, responsive role. Since scientists focus on fine-grained modifications, AI that does not fully understand the precise context may interfere with their work. 

Making users’ insight states transparent may help HCI researchers design AI roles in scientific discovery interfaces, aligning with the core principle of human-AI collaboration, where humans retain initiative over core tasks~\cite{wang2019designing, amershi2019guidelines}.
For example, if users share their experiences of abductive leaps with the AI during a session, the AI can reduce its intrusiveness in later phases by aligning with user input when needed, thereby facilitating human-AI interaction.

\subsection{Limitation and Future Work}

First, evaluating the quality of compounds generated by \system solely using model-based scores and counts is inherently limited. Definitive assessment requires scientific validation through wet-lab experiments. However, such experiments are costly, with approximately \$2,500 per compound, making large-scale validation impractical within the scope of this study. In future work, we plan to integrate experimentally validated datasets as ground truth, enabling more rigorous evaluation in lead optimization with \system.

Second, both the duration of the user study (30 minutes) and the size of the participant pool (N = 10) limit the generalizability of our findings. While participants were able to discover several high-quality compound candidates in a short timeframe, in real-world practice, identifying compounds typically takes at least a week. Moreover, although ten participants completed expert studies, the sample may not fully represent the diversity of the drug discovery domain.
For future work, by collaborating with more diverse drug discovery labs, we aim to conduct long-term user studies (at least one week) to enhance the generalizability of \system.

Third, regarding the analysis of the abductive leap, we aim to extend our analysis beyond the experimental group to the control group by instrumenting both conditions with a comparable insight-capture mechanism for a more detailed comparison of abductive leap patterns across conditions.

\section{Conclusion}
In this paper, we present the Co-abduction framework, a novel approach to human-AI collaboration for abductive reasoning and the generation of new, promising hypotheses. To operationalize this framework, we developed \system, which supports co-abductive reasoning through three stages in ligand optimization: 1) efficient observation, 2) systematic strategy identification, and 3) coherent multi-strategy synthesis.
Results from a study involving 10 domain experts demonstrate that \system facilitates abductive reasoning and supports the generation of higher-quality, more diverse hypotheses, demonstrating its applicability to scientific domains that require novel and promising hypotheses, including materials science, biology, and chemistry.
Our work offers a new perspective on supporting hypothesis generation, emphasizing the importance of abductive reasoning in human-AI collaboration.


\clearpage
\bibliographystyle{ACM-Reference-Format}
\bibliography{Reference}

\clearpage
\appendix

\section{Appendix}

\subsection{Molecule Generation Pipeline}\label{pipeline}

\begin{figure}[H]
    \centering
    \includegraphics[width=1.0\linewidth]{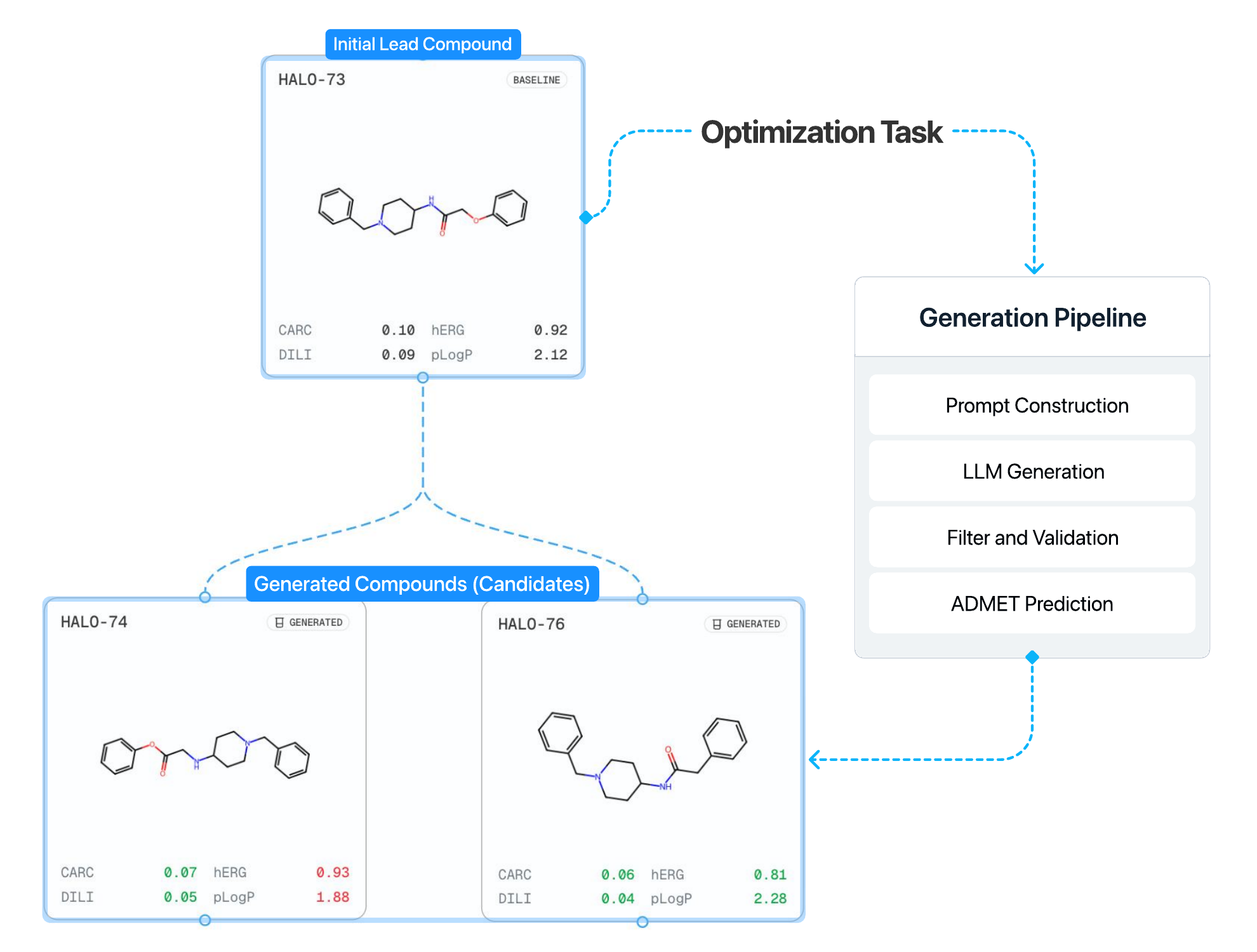}
    \caption{Overview of the compound generation pipeline. Users select an optimization task, which is used to construct a prompt for Llama~3.1~8B-Instruct fine-tuned on C-MuMOInstruct~\cite{dey2025gellm3o}. Generated compounds are filtered and scored before display.}
    \label{fig:llm_pipeline}
\end{figure}

Figure~\ref{fig:llm_pipeline} illustrates the molecule generation pipeline. Users select a source molecule and choose an optimization task from predefined profiles that group ADMET properties by therapeutic relevance. Users can modify the molecular structure and review physicochemical properties in a web-based chemical editor before initiating generation.

\begin{figure*}
    \centering
    \includegraphics[width=1\linewidth]{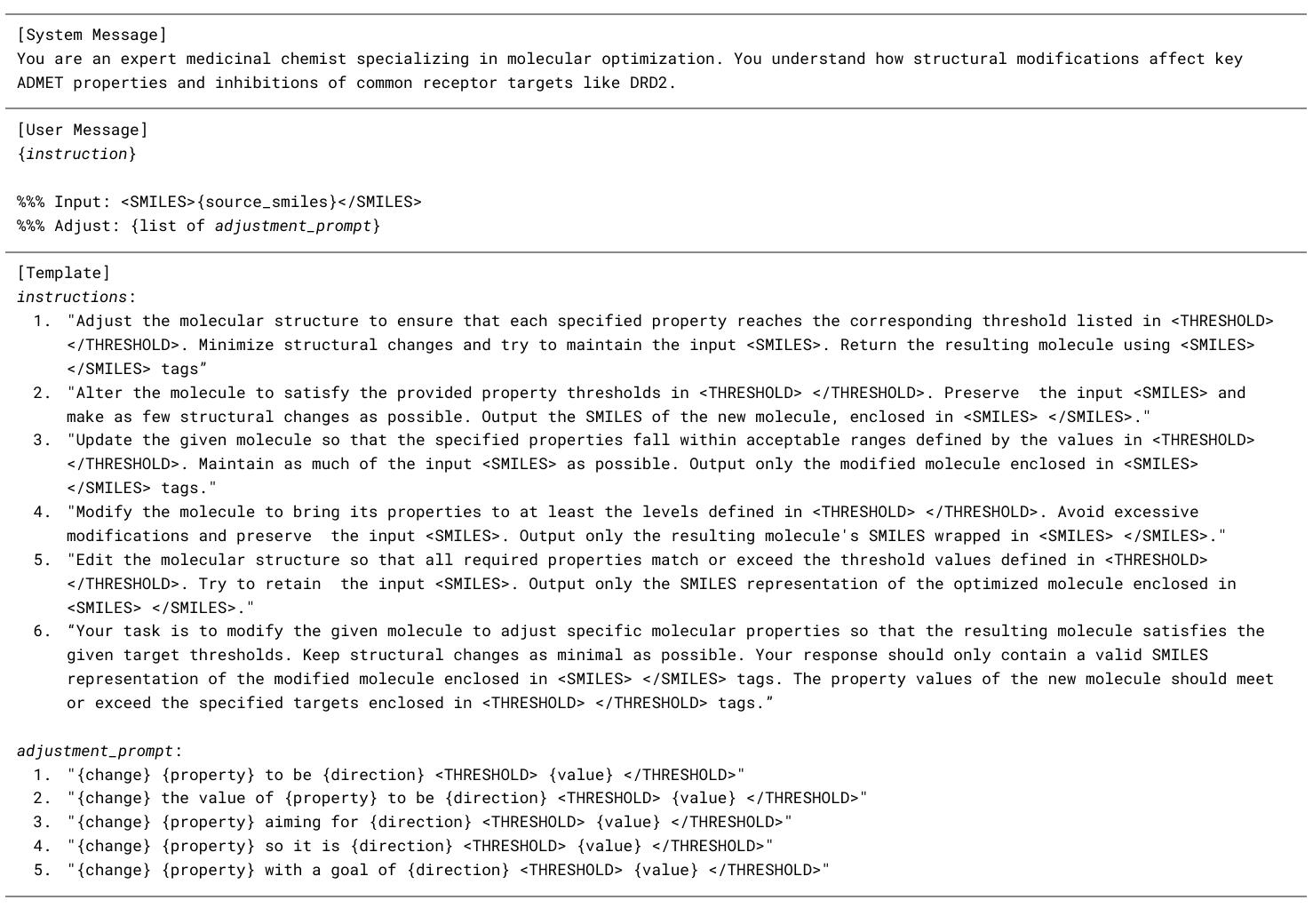}
    \caption{The prompts used to generate molecules through Llama~3.1~8B-Instruct. \textit{Source\_smiles} refers to the SMILES string combining the core scaffold and preserving R-groups. \textit{Instruction} is selected at random from the “instructions” list in Template. For each optimizing property, an \textit{adjustment\_prompt} is appended by randomly selecting a prompt from the “adjustment\_prompt” list in the Template, with variables (denoted by \{ \}) filled in for each ADMET property.}
    \label{fig:gen_prompt}
\end{figure*}

The system constructs a structured prompt encoding the source molecule as a SMILES string alongside the task's property targets and optimization directions (Figure~\ref{fig:gen_prompt}). The prompt is processed by Llama~3.1~8B-Instruct fine-tuned on C-MuMOInstruct~\cite{dey2025gellm3o}, an instruction-tuning dataset for controllable multi-property molecule optimization. The model generates multiple candidate SMILES per request. Output SMILES are validated, canonicalized, and deduplicated via RDKit.

Each candidate undergoes a maximum common substructure (MCS) filter to ensure scaffold preservation. The MCS between candidate and source is computed using RDKit's FMCS algorithm with whole-ring and ring-bond-type constraints. Candidates whose MCS covers less than 60\% of the source's heavy atoms are discarded:
\[
\text{retained} = \left\{ m \in \text{candidates} \;\middle|\;
\frac{|\text{MCS}(m, \text{source})|}{|\text{source}|} \geq 0.6 \right\}
\]

Retained candidates are scored using ADMET-AI~\cite{swanson2024admet} (carcinogenicity, hERG, DILI) and Therapeutic Data Commons oracles (DRD2, pLogP), alongside RDKit-computed physicochemical descriptors and drug-likeness metrics. Each candidate is rendered on the canvas as a child node linked to its source by a generation edge.

\begin{table*}
  \centering
  \begin{threeparttable}
    \caption{Computational evaluation of compound generation across scaffold types and tasks.}
    \label{tab:compound-generation}
    \begin{tabular}{lcccccc}
      \toprule
      & \multicolumn{2}{c}{Any Property Success (\%)} & \multicolumn{2}{c}{Hallucination Rate (\%)} & \multicolumn{2}{c}{Scaffold Preservation (\%)} \\
      \cmidrule(lr){2-3} \cmidrule(lr){4-5} \cmidrule(lr){6-7}
      Scaffold Type & Task A & Task B & Task A & Task B & Task A & Task B \\
      \midrule
      Bemis-Murcko (murcko)\tnote{a}
        & 72.3 & 100.0 & 0.0 & 38.3 & 89.8 & 100.0 \\
      Ring System (ring only)\tnote{b}
        & 79.3 & 100.0 & 0.0 & 29.7 & 93.8 & 100.0 \\
      Exact Structure (full molecule)\tnote{c}
        & 73.8 & 100.0 & 0.0 & 39.8 & 90.6 & 100.0 \\
      \bottomrule
    \end{tabular}
    \begin{tablenotes}
      \item[a] Bemis-Murcko scaffold: the core ring structure with linkers, removing side chains.
      \item[b] Ring system: only ring atoms are preserved.
      \item[c] Exact structure: the complete molecular structure.
    \end{tablenotes}
  \end{threeparttable}
\end{table*}

\begin{figure}[H]
  \centering
  \includegraphics[width=0.95\columnwidth]{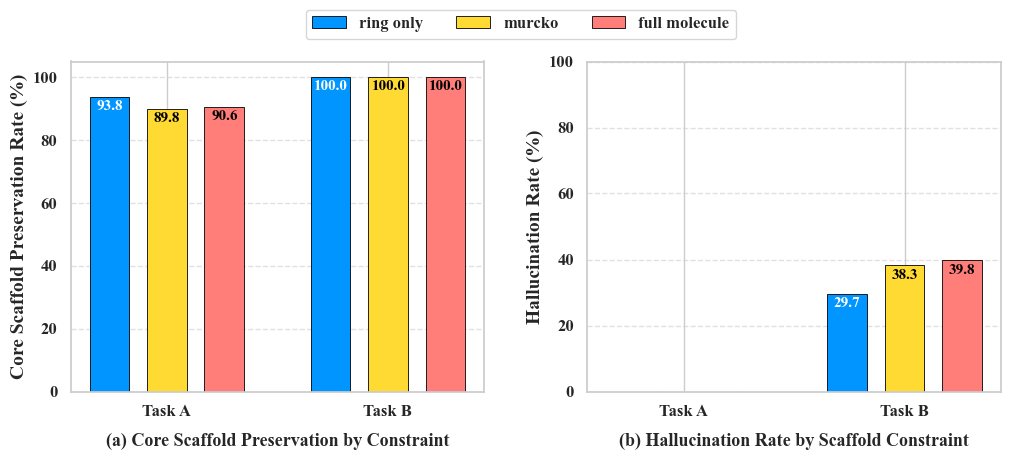}
  \caption{(a) Core scaffold preservation rate across constraint types. (b) Hallucination rate measured as RDKit parsing failures. Task~A represents out-of-distribution optimization (CARC, hERG, LIVER, PLogP). Task~B represents in-distribution optimization (DRD2, PLogP). $n{=}1{,}024$ candidates per task.}
  \label{fig:scaffold}
\end{figure}

\begin{figure}[H]
  \centering
  \includegraphics[width=0.95\columnwidth]{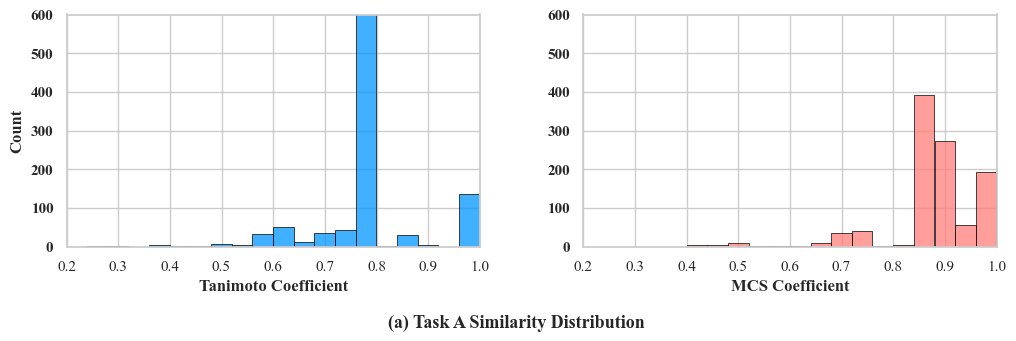}
  \includegraphics[width=0.95\columnwidth]{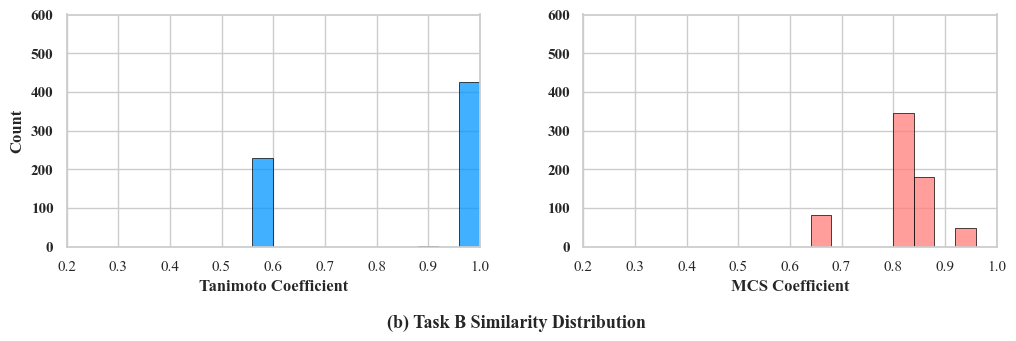}
  \caption{Structural similarity distributions between seed and generated molecules. (a) Task~A shows tight distributions indicating conservative modifications. (b) Task~B exhibits bimodal Tanimoto coefficients suggesting variable generation behavior.}
  \label{fig:task-sim}
\end{figure}
\subsection{Computational Evaluation of Molecule Generation}\label{compoundstudy}

We evaluated the generation reliability of Llama~3.1~8B-Instruct fine-tuned on C-MuMOInstruct~\cite{dey2025gellm3o} using property combinations from our user study tasks. Task~A optimized CARC, hERG, DILI, and pLogP, a combination absent from training data (out-of-distribution). Task~B optimized DRD2, DILI, pLogP, and QED, matching the training distribution (in-distribution). We generated 1,024 candidates per task and measured hallucination rate (RDKit parsing failures), scaffold preservation, and any-property success across three scaffold constraints: Bemis-Murcko, ring system, and complete molecular structure.

Task~A produced 0\% hallucination, 89.8--93.8\% scaffold preservation, and 72.3--79.3\% any-property success. Task~B achieved 100\% scaffold preservation and 100\% any-property success but exhibited 29.7--39.8\% hallucination (Figure~\ref{fig:scaffold}, Table~\ref{tab:compound-generation}). The out-of-distribution task outperformed on validity. We attribute this to learned biases for DRD2 optimization in the training data. When scaffold constraints conflict with these patterns, the model forces modifications that produce invalid molecules. Structural similarity confirms this: Task~A generated conservative modifications (Tanimoto centered at 0.8), whereas Task~B showed bimodal behavior with peaks at 0.6 and 1.0 (Figure~\ref{fig:task-sim}).

\subsection{Framework for Evaluating Properties}\label{tdetail}

The evaluated property scores are grouped into three categories: ADMET properties, physicochemical properties, and Drug-Likeness. To maintain a clean interface, each category is displayed within an expandable accordion section. Within each category, the properties scores are listed in a structured tabular format. In addition to properties scores, similar compounds for each molecule, retrieved through PubChem's PUG REST \footnote{https://pubchem.ncbi.nlm.nih.gov/docs/pug-rest}.

\subsection{Details on properties}\label{property}

We used a total of five properties in Tasks A and B. Task A included PlogP, hERG, CARC, and LIVER, while Task B included PlogP, QED, DRD2, and LIVER.

\begin{itemize}
    \item PlogP (Penalized logP)~\cite{hoffman2022optimizing}: PlogP is the octanol–water partition coefficient (logP); higher values are preferred. 
    \item hERG~\cite{karim2021cardiotox}: hERG refers to whether a compound blocks a heart potassium channel (called hERG); lower values are preferred.
    \item CARC (Carcinogenicity)~\cite{van2016critical}: CARC refers to the risk that a compound could cause cancer; lower values are preferred.
    \item DRD2~\cite{jandaghi2016expression}: DRD2 refers to the probability that a compound blocks the dopamine D2 receptor (DRD2); higher values are preferred.
    \item QED~\cite{bickerton2012quantifying}: QED refers to overall drug-likeness by incorporating molecular weight, lipophilicity, and hydrogen bonding ability– higher QED is desired for better drug-likeness
    \item LIVER (Drug‑induced liver injury risk)~\cite{hosack2023drug}: LIVER risk describes the propensity of a compound to cause liver injury; lower values are preferred.
\end{itemize}

\subsection{Baseline Interface}\label{baselinedesign}
\begin{figure*}
    \centering
    \includegraphics[width=1\linewidth]{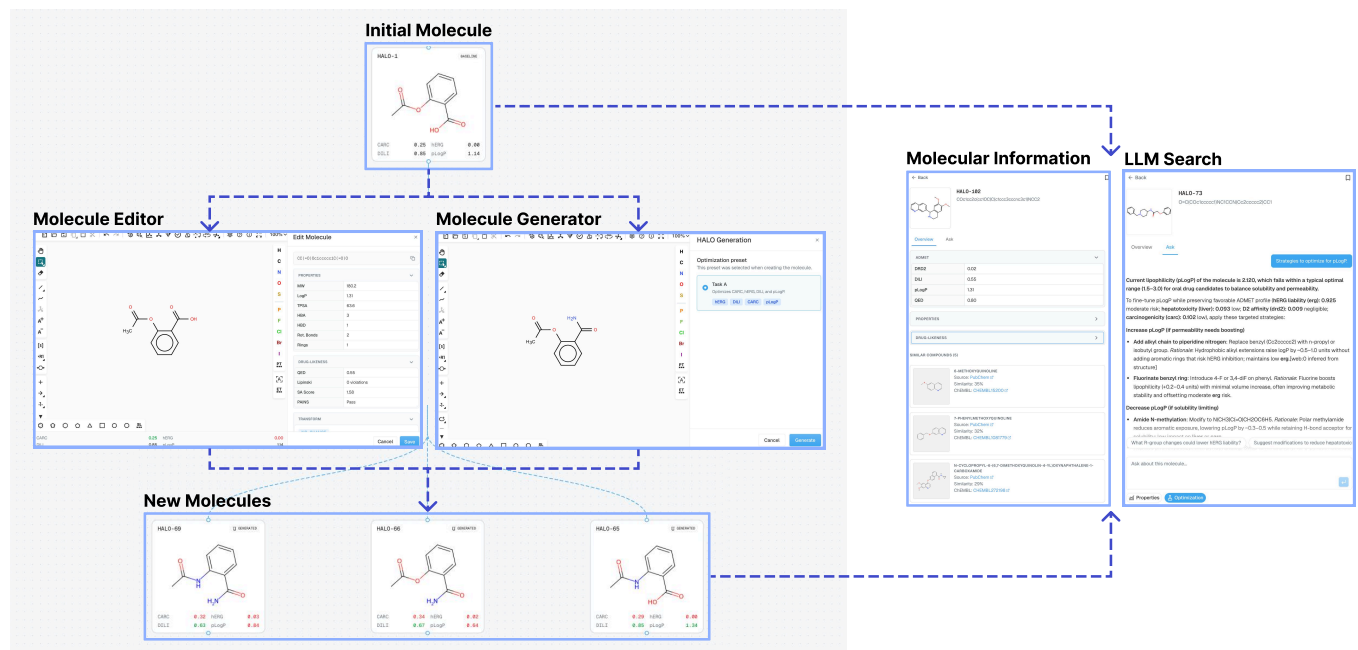}
    \caption{The baseline interface used for user studies. The Molecule Editor and Molecule Generator appear as pop-up windows when the user selects from a context menu. Molecular Information and LLM appear in the side panel when expanded. (\textcolor{improved_green}{green: augmented} and \textcolor{worsened_red}{red: worsened})}
    \label{fig:control}
\end{figure*}

To isolate and evaluate co-abductive reasoning, we developed a baseline system (Figure \ref{fig:control}). The baseline includes non-co-abductive components and excludes three co-abductive features (e.g., \fone, \ftwo, and \fthree).
First, users can use the generative AI model in \system to generate molecules on the trajectory map, and then check the property scores for four targeted properties (\textcolor{improved_green}{green: augmented} and \textcolor{worsened_red}{red: worsened}). Newly edited or generated molecules appear beneath the original molecule, connected by a dashed line.  
If users choose to edit a molecule, the pop-up window is divided into a chemical editor and a properties panel that updates in real time as changes are made in the editor on the right. 
By right-clicking on any molecule node, users can access a side panel where the properties and LLM interface are located. The properties display includes various ADMET, physicochemical, and Drug-Likeness scores, as well as similar compounds found in PubChem. The LLM interface allows users to input queries to assist with their tasks (e.g., searching for references, strategies for optimization, etc.).

\begin{figure*}[h]
    \centering
    \includegraphics[width=1.0\textwidth]{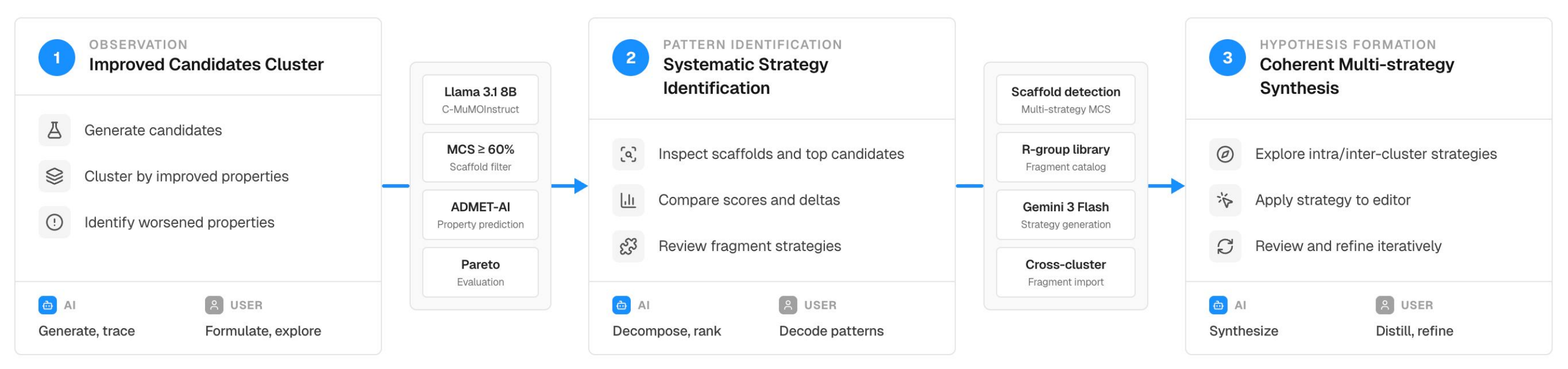}
    \caption{Overview of the HALO pipeline. A three-stage workflow: (1) Observation, generate and cluster molecules by properties; (2) Strategy Identification, analyze scaffolds, scores, and fragment strategies; (3) Strategy Synthesis, combine and apply strategies to iteratively refine candidates.}
    \label{fig:halo_pipeline}
\end{figure*}

\subsection{Technical Details of \system{}}\label{techdetails}

We describe the technical details of \system{} in terms of three main components: \fone, \ftwo, and \fthree{} (Figure~\ref{fig:halo_pipeline}).

\subsubsection{Stage 1: Candidate Generation Pipeline}

The generation pipeline converts a user's optimization task into scored candidate molecules. The system constructs a structured prompt that encodes the source molecule as a SMILES string, selected ADMET property targets, and their optimization directions (e.g., decrease hERG, increase pLogP). Each property direction and threshold is formatted as an adjustment instruction.

The prompt is processed by Llama~3.1~8B-Instruct, fine-tuned on C-MuMOInstruct~\cite{dey2025gellm3o}. The model generates multiple SMILES candidates per request, which are extracted using \texttt{<SMILES>}\ldots\texttt{</SMILES>} delimiters, then validated, canonicalized, and deduplicated with RDKit.

\subsubsection{Stage 2: Cluster Annotation Pipeline}

The annotation pipeline runs when a user opens a cluster's detail view, computing five parallel analyses that characterize the cluster's structure-activity profile.

\paragraph{Core scaffold detection.}
Shared scaffolds are identified through a two-phase MCS search. First, a global MCS is computed across all cluster members. To address diverse clusters where the global MCS may be trivially small, a per-scaffold MCS is also computed within the five largest Murcko scaffold subgroups. Candidate cores are filtered by coverage, minimum heavy atom count (at least 60\% of the median scaffold size), and decomposition quality. A greedy set-cover algorithm selects up to three complementary cores that maximize cluster coverage.

\paragraph{R-group library construction.}
Each selected core defines a decomposition using RDKit's R-group decomposition. Matching molecules are split into the shared scaffold and variable side chains at numbered attachment sites, producing a frequency-ranked catalog of observed R-group fragments at each site.

\paragraph{Chemotype profiling.}
Each molecule is annotated for ring systems, functional groups, and structural alerts using medicinal chemistry catalogs. Per-molecule profiles are aggregated at the cluster level into motif frequencies and alert themes.

\paragraph{Cluster overview generation.}
Pareto statistics, property ranges, scaffold distributions, and alert themes are assembled into structured context and passed to Gemini~3~Flash, which generates a natural-language summary of the cluster's strengths and weaknesses.

\subsubsection{Stage 3: Strategy Synthesis Pipeline}

The strategy pipeline transforms cluster annotations into actionable optimization recommendations by integrating inter-cluster and intra-cluster structural insights.

\paragraph{Source molecule selection.}
Candidates are ranked using Pareto evaluation criteria (objectives met, Pareto optimality, baseline improvement), and the top three are chosen as source molecules. Each source is decomposed relative to the cluster's core to show its current R-group assignment and available alternatives from the R-group library.

\paragraph{Cross-cluster fragment import.}
The system evaluates neighboring clusters by running abbreviated Pareto evaluations and selecting the highest-ranked molecules. Pairwise MCS computations between primary sources and neighboring molecules identify regions of structural overlap. Where sufficient overlap exists, the system extracts R-group fragments from neighbors that are absent from the primary cluster's library, introducing structurally validated modification options from related chemical contexts.

\paragraph{Strategy generation.}
A structured prompt containing the source molecules, their R-group assignments, all available intra-cluster and inter-cluster alternatives, and the optimization objectives is processed by Gemini~3~Flash. Each returned strategy specifies a title, target property, natural-language rationale, source molecule, attachment site, current fragment SMILES, and replacement fragment SMILES. Strategies are classified as intra-cluster (from the cluster's own library) or inter-cluster (from neighboring clusters).

\paragraph{Post-processing and validation.}
All SMILES strings are canonicalized and validated. Common LLM encoding errors in aromatic fragments are corrected using standard ring templates.

\begin{figure*}
\centering
  \includegraphics[width=1.8\columnwidth]{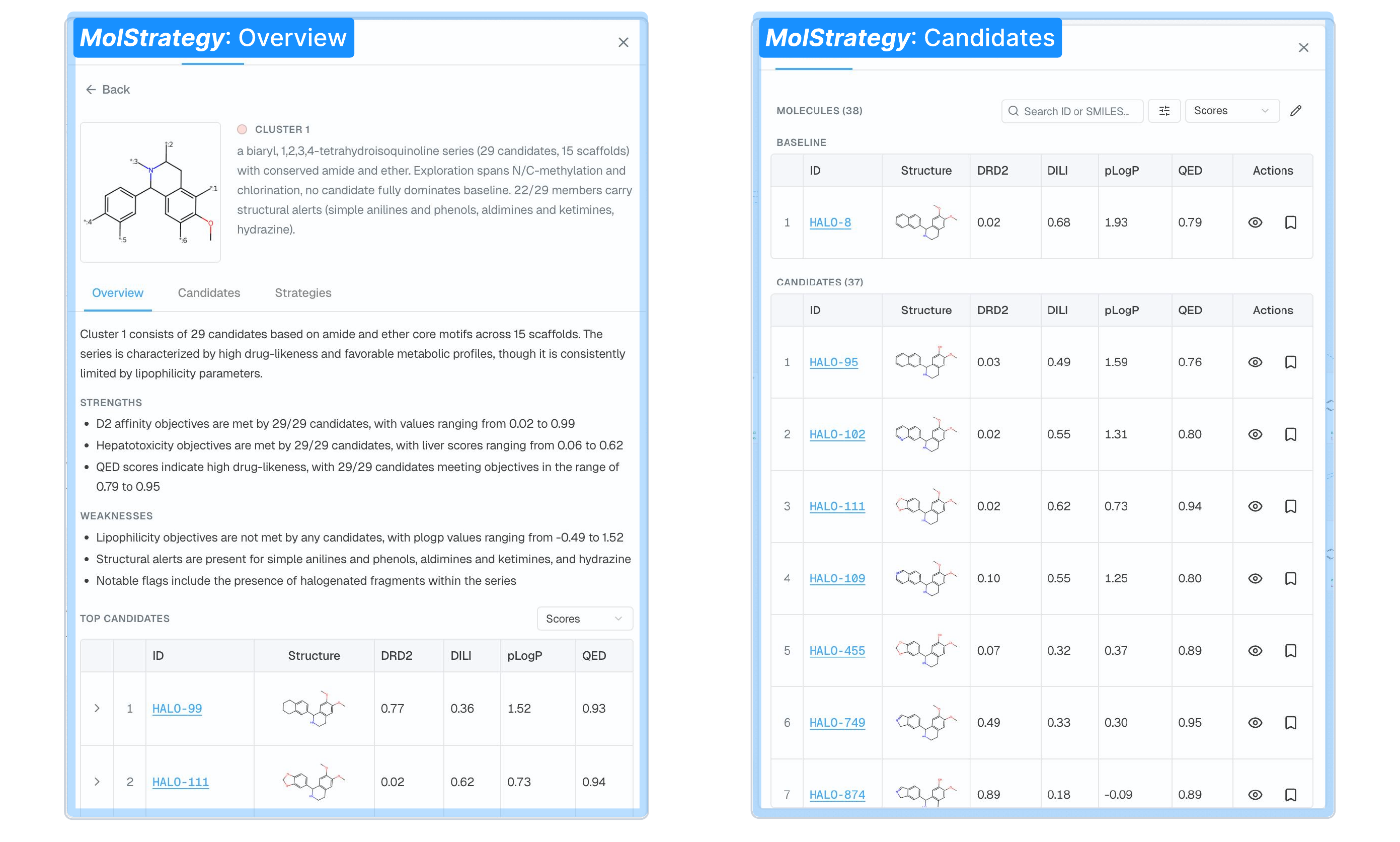}
  \caption{
  Two tabs of \ftwo. The overview tab displays an AI-generated summary of the cluster alongside its identified strengths and weaknesses. A scaffold image shows the shared core structure identified across cluster members, and a top candidates section highlights the highest-ranked molecules based on Pareto evaluation. The candidates tab lists all cluster members in a sortable table, where users can toggle between absolute property scores and improvement deltas relative to the baseline. Delta values are color-coded to indicate improvements and regressions.}
  ~\label{fig:mol_sys}
\end{figure*}

\begin{figure*}
\centering
  \includegraphics[width=1.8\columnwidth]{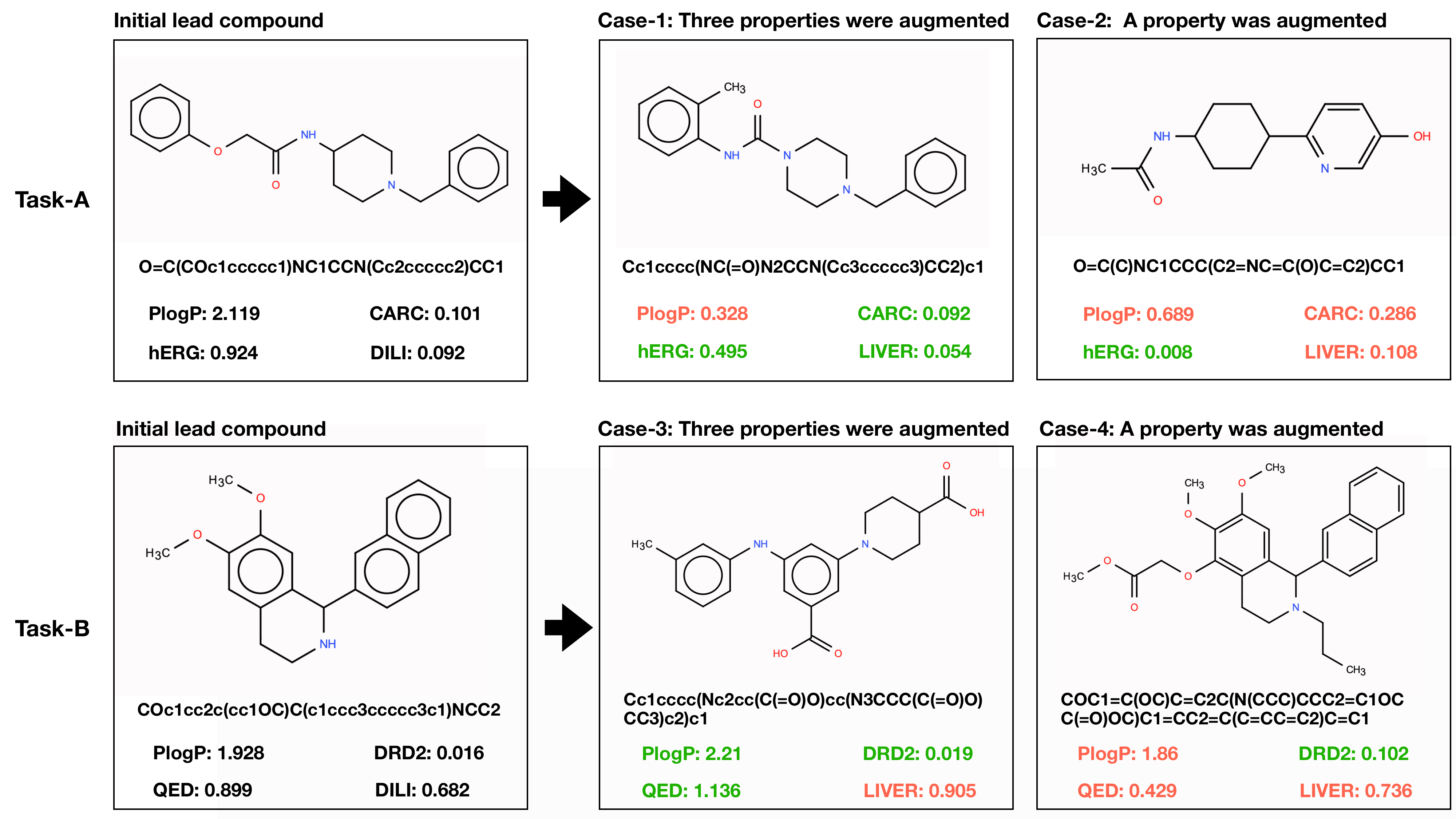}

  \caption{The initial lead compounds and the desired properties of Task-A and Task-B. Four representative cases illustrate contrasting outcomes: Cases 1 and 3 show ideal results, where three properties were improved simultaneously, whereas Cases 2 and 4 show less desirable results, where only one property was improved at the expense of the others (\textcolor{improved_green}{green: augmented} and \textcolor{worsened_red}{red: worsened}). For PlogP, DRD2, and QED, higher values indicate more desirable outcomes, whereas for hERG, CARC, and LIVER, lower values are preferred. Please refer to the details of these properties {\S}\ref{property} in the appendix.}
  ~\label{fig:task_ex}
\end{figure*}

\begin{table*}[t]
    \centering
    \caption{Definitions used to label participants' actions for the user studies.}
    \renewcommand{\arraystretch}{1.25}
    \begin{tabular}{c|>{\raggedright\arraybackslash}p{12cm}}
    \toprule
    \textbf{Label} & \textbf{Description} \\
    \midrule
    
    Molecule Generated (AI) &
    \textbullet\ Participant generates from a preexisting molecule. \\

    \midrule
    
    \multirow{2}{*}{Molecule Generated (Manual)} &
    \textbullet\ Participant saves a manually edited molecule. \\
    
    &\textbullet\ Participant edits a preexisting molecule before generating. \\

    \midrule
    
    \multirow{3}{*}{\fone} &
    \textbullet\ Participant modifies the filters and applies changes. \\
    
    & \textbullet\ Participant utilizes \fone panel to select or deselect at least one property from a cluster. \\
    
    & \textbullet\ Participant utilizes selection tool to group or ungroup at least two molecules.\\

    \midrule
    
    \ftwo &
    \textbullet\ Participant actively interacts (e.g., scrolls or clicks) with \ftwo panel. \\

    \midrule
    
    \fthree &
    \textbullet\ Participant clicks on at least one strategy in \fthree panel. \\

    \midrule
    
    Insight &
    \textbullet\ Participant clicks on the “Insight” button. \\
    
    \bottomrule
    \end{tabular}
    \label{tab:user_log_definition}
\end{table*}

\begin{table*}[t]
\centering
\caption{Participant demographics and study participation. Senior: Senior Researcher; Postdoc: Postdoctoral Researcher; PhD Cand.: Ph.D. Candidate.}
\renewcommand{\arraystretch}{1.15}

\begin{tabular}{c c c c c c}
\toprule
\textbf{ID} & \textbf{Age} & \textbf{Profession} & \textbf{YoE} & \textbf{Formative} & \textbf{User Study} \\
\midrule
$P{1}$  & 36--45 & Senior R.  & 18 & $\checkmark$ &              \\
$P{2}$  & 26--35 & PhD Cand.  & 4  & $\checkmark$ &              \\
$P{3}$  & 36--45 & Postdoc    & 10 & $\checkmark$ & $\checkmark$ \\
$P{4}$  & 36--45 & Senior R.  & 14 & $\checkmark$ & $\checkmark$ \\
$P{5}$  & 26--35 & Senior R.  & 15 & $\checkmark$ & $\checkmark$ \\
$P{6}$  & 26--35 & Postdoc    & 9  &              & $\checkmark$ \\
$P{7}$  & 26--35 & Postdoc    & 10 &              & $\checkmark$ \\
$P{8}$  & 26--35 & Postdoc    & 10 &              & $\checkmark$ \\
$P{9}$  & 36--45 & Senior R.  & 13 &              & $\checkmark$ \\
$P{10}$ & 26--35 & Postdoc    & 8  &              & $\checkmark$ \\
$P{11}$  & 26--35 & Postdoc   & 10 &              & $\checkmark$ \\
$P{12}$ & 26--35 & Postdoc    & 10  &              & $\checkmark$ \\
\bottomrule
\end{tabular}
\label{tab:participant}
\end{table*}

\end{document}